\newif\ifpr

\ifpr
\documentclass[amsmath, longbibliography, amssymb, aps, prl, superscriptaddress, twocolumn, floatfix, 10pt]{revtex4-1} 

\else
\documentclass[amsmath, longbibliography, amssymb, aps, prl, superscriptaddress, twocolumn, floatfix, 10pt, nofootinbib]{revtex4-1} 

\fi
\usepackage[utf8]{inputenc}
\usepackage{textgreek} 

\usepackage[bbgreekl]{mathbbol}
\usepackage{amsfonts}
\DeclareSymbolFontAlphabet{\mathbb}{AMSb}
\DeclareSymbolFontAlphabet{\mathbbl}{bbold}

\usepackage{gensymb,amssymb,graphicx,color,amsmath,mathtools,bm,multirow}
\usepackage{hyperref}
\usepackage[normalem]{ulem}
\usepackage[caption=false]{subfig} 
\usepackage{braket}

\newcommand{\secref}[1]{Sec.\,\ref{#1}}
\newcommand{\refcite}[1]{Ref.\,\cite{#1}}
\newcommand{\refscite}[1]{Refs.\,\cite{#1}}
\newcommand{\eqnref}[1]{Eq.\,\eqref{#1}}
\newcommand{\eqsref}[1]{Eqs.\,\eqref{#1}}
\newcommand{\figref}[1]{Fig.\,\ref{#1}}

\newcommand{\sfigref}[2]{Fig.\,\hyperref[#1]{\ref{#1}(#2)}}
\newcommand{\tabref}[1]{Tab.\,\ref{#1}}

\newcommand{\appref}[1]{Appendix\,\ref{#1}}

\definecolor{kspink}{RGB}{200,0,200}

\newcommand{\xx}{\mathbbl{x}}

\DeclareMathOperator*{\Otimes}{\text{\scalebox{1.2}{$\otimes$}}}
\DeclareMathOperator{\RE}{Re}
\DeclareMathOperator{\IM}{Im}
\DeclareMathOperator*{\Exp}{\text{\scalebox{1.2}{$\mathbb{E}$}}}
\DeclareMathOperator*{\tExp}{\mathbb{E}}

\hypersetup{
  colorlinks = true,
  urlcolor = blue,
  pdfauthor = {Kevin Slagle and John Preskill},
  pdftitle = {Slagle and Preskill - 2023 - Emergent Quantum Mechanics at the Boundary of a Local Classical Lattice Model}
}

\newcommand{\EmQM}{E\texorpdfstring{\MakeLowercase{m}}{m}QM}

\begin{document}

\title{Emergent Quantum Mechanics at the Boundary of a Local Classical Lattice Model}

\author{Kevin Slagle}
\affiliation{Department of Electrical and Computer Engineering, Rice University, Houston, TX 77005 USA}
\affiliation{Department of Physics, California Institute of Technology, Pasadena, California 91125, USA}
\affiliation{Institute for Quantum Information and Matter and Walter Burke Institute for Theoretical Physics, California Institute of Technology, Pasadena, California 91125, USA}

\author{John Preskill}
\affiliation{Institute for Quantum Information and Matter and Walter Burke Institute for Theoretical Physics, California Institute of Technology, Pasadena, California 91125, USA}
\affiliation{AWS Center for Quantum Computing, Pasadena, California 91125, USA}

\date{\today}

\begin{abstract}
We formulate a conceptually new model in which quantum mechanics emerges from classical mechanics.
Given a local Hamiltonian $H$ acting on $n$ qubits,
  we define a local classical model with an additional spatial dimension
  whose boundary dynamics is approximately---but to arbitrary precision---described by Schr\"{o}dinger's equation and $H$.
The bulk consists of a lattice of classical bits
  that propagate towards the boundary through a circuit of stochastic matrices.
The bits reaching the boundary are governed by a probability distribution
  whose deviation from the uniform distribution can be interpreted as the quantum-mechanical wavefunction.
Bell nonlocality is achieved because information can move through the bulk much faster than the boundary speed of light.
We analytically estimate how much the model deviates from quantum mechanics, and we validate these estimates using computer simulations.
\end{abstract}

\maketitle

\section{Introduction}
General relativity and the Standard Model of particle physics are not exact descriptions of reality;
  rather they emerge as low-energy effective field theory descriptions of some underlying theory (e.g. string theory).
A characteristic signature of emergence at low energy or long time and distance scales
  is that the resulting physics is typically well-described by remarkably simple equations,
  which are often linear (e.g. the harmonic oscillator)
  or only consist of lowest-order terms in an effective Lagrangian.

In principle, it is possible that quantum mechanics is also only an approximate description of reality.
Indeed, Schr\"{o}dinger's equation is a simple linear differential equation, suggesting that it might arise as the leading approximation to a more complete model.
The advent of quantum computing opens opportunities to probe and test quantum mechanics in an unprecedented regime. Much evidence indicates that if standard quantum theory is exactly correct, then the cost of simulating a quantum computation with a classical computer must grow exponentially with the size of the quantum computer \cite{AaronsonSupremacy}. To the extent possible, this extraordinary hypothesis about the quantum world should be tested in the laboratory.
Indeed, if quantum theory actually emerges from an underlying classical model, then this exponential scaling must eventually fail for real devices.
Therefore, aside from verifying Bell nonlocality  \cite{BellNonlocality} and studying the behavior of macroscopic superpositions \cite{superpositions}, we should also conceive and perform experiments that characterize the computational power of nature \cite{FalsifiableQM,SlagleTesting,HooftBook}.

Numerous experiments \cite{experimentRev1992,testQM2000,g2Review,photonTests,BellNonlocality,GoogleSupremacy,ZuchongzhiAdvantage2,PhotonAdvantage} and theoretical observations \cite{RudolphStateReality,PolchinskiNonlinearEPR,GisinNonlinearEPR,PostIQP,AaronsonPsiEpistemic,Mielnik} significantly constrain, but do not completely rule out, possible deviations from standard quantum theory. For example, measurements of the anomalous magnetic dipole moment of the electron \cite{electronMoment} agree with quantum predictions up to roughly ten digits of precision. However, these experiments, and most other current tests of quantum theory, probe properties of matter with relatively low computational complexity, and so might be insensitive to deviations from quantum theory that become evident only for more complex states \cite{AaronsonSureShor}. Quantum theory also successfully predicts properties of ground states and of low-energy dynamics for many materials and molecules, but here too the detailed agreement between theory and experiment has mostly been limited to quantum states that are not profoundly entangled \cite{GarnetPolynomial,bubbleCollisions}, and so the successful predictions do not rule out departures from quantum predictions for states of high complexity. To probe the high-complexity regime convincingly, highly excited matter should be carefully studied. It may also be necessary to measure many observables, since classically tractable models of thermalization \cite{DeutschETH,SrednickiETH} and emergent hydrodynamics \cite{DMT,AltmanThermalizationMPS,DAOE} may suffice for explaining the observed data when only a few degrees of freedom are measured. In contrast to more conventional experimental tools, future quantum computers that prepare highly entangled states and perform intricate measurements will be well equipped for probing the behavior of matter in the regime far beyond the reach of efficient classical simulation \cite{GoogleSupremacy,ZuchongzhiAdvantage2,PhotonAdvantage}.

Though models in which quantum dynamics emerges from underlying classical dynamics should be testable in the high complexity regime, such tests need not be applicable to other proposed modifications of standard quantum theory. For example, in models with intrinsic wavefunction collapse \cite{collapseReview}, quantum error correction \cite{errorCorrectionIntro} might overcome the damaging effects of the intrinsic noise, restoring the full computational power of quantum theory. Furthermore, generic non-linear corrections to Schr\"odinger's equation may well enhance rather than diminish the computational power of quantum systems \cite{AbramsNonlinearNP}, while our goal is to explore whether nature could be computationally weaker, not stronger, than standard quantum theory predicts.

Such considerations motivate the quest for testable models in which quantum mechanics emerges from classical mechanics and for which deviations from standard quantum theory are detectable in the high-complexity regime \cite{Hooft,HooftBook,Adler,VanchurinEntropic,VanchurinWorld,VanchurinNeural,Nelson2012,Wolfram2020,InteractingWorlds,Palmer,Torrome}.
In pursuit of this quest, we aim to construct a local classical lattice model that exhibits emergent quantum mechanics (EmQM) \cite{EmQM}.
We define a local classical model to consist of a lattice,
  where the state at each lattice site is defined by a finite list of numbers (which does not grow with the system size),
  and the time evolution of each lattice site only depends on the state of nearby sites.
The time evolution is allowed to be stochastic and either continuous or discrete.
Cellular automata and local classical lattice Hamiltonians (and Lagrangians)
  are examples of local classical models.

We consider a model to exhibit EmQM if its slowly-varying and long-distance physics is well-described by Schr\"{o}dinger's equation:\footnote{%
  The ``$\cdot$'' denotes matrix and vector dot products.
  We avoid bra-ket notation since we will relate the wavefunction to a classical probability vector $\bm{P}$, and we do not want to place $\bm{P}$ in a ket or mix notation styles.}
\begin{equation}
  \partial_t \Psi = -i H \cdot \Psi \label{eq:Schro}.
\end{equation}
$\Psi$ is the wavefunction, which encodes the state of the system,
  while $H$ is the Hamiltonian\footnote{
    In this work, we will only consider local Hamiltonians for a lattice of qubits,
      for which $H$ is a Hermitian matrix and $\Psi$ is a complex-valued vector.
    A quantum Hamiltonian is local if it is a sum of terms that each act only on nearby qubits.},
  which defines the dynamics.

In a sense, the classical model is performing an approximate simulation of suitably encoded Schr\"{o}dinger dynamics. We want more than just that --- to the extent possible we want the classical dynamics to be a reasonable model of how nature might really behave.  Typical classical simulation algorithms will not fit the bill, for example because the classical dynamics is spatially non-local or because the required number of local degrees of freedom increases exponentially with system size (e.g. tensor network methods \cite{OrusTN,IsoTNDynamics} require an exponentially large bond dimension\footnote{Spatial locality is also a challenge for tensor networks, although a spatially local algorithm has been derived in one dimension. \cite{ParallelDMRG}}). Drawing inspiration from neutral network algorithms \cite{NTK,MendlNeural,SchmittNeural2D,PollmannNeural,JonssonNN}, we seek models without such shortcomings. 

We partially succeed in the following sense.
Given any local Hamiltonian and initial value wavefunction for $n$ qubits in $D$ spatial dimensions,
  we can define a local classical lattice model in $D+1$ spatial dimensions
  whose $D$-dimensional boundary dynamics can be well-approximated by Schr\"{o}dinger's equation if
  the extra spatial dimension has a length $S$ that is exponentially large in $n$, i.e. if $S \gg 2^n$.

We view this as only a partial success because if $S \gg 2^n$ does not hold,
  then the boundary dynamics instead obeys a Schr\"{o}dinger's equation with a highly nonlocal Hamiltonian
  (i.e. its terms are geometrically nonlocal and also act on many qubits at once).
In order to be a model of EmQM in our universe,
  we would like to view $n$ as the number of (possibly Planckian-sized) qubits in our universe
  (with an effective low-energy Lagranian or Hamiltonian consistent with the standard model \cite{LevinWenEmergentPhotons,WenAnomalies}).
More conservatively, we would like to take $n$ to at least be as large as the number of qubits needed to describe a macroscopic region of space,
  e.g. certainly larger than Avogadro's number: $n > 10^{23}$.
Thus, in order to be consistent with local quantum dynamics,
  the extra dimension would have to be tremendously long, e.g. $S \gg 2^{10^{23}}$.
Future work is necessary to determine if it is possible to alleviate the $S \gg 2^n$ requirement.
For example, one might instead demand that an
   EmQM model with fixed $S \gg 2^{\widetilde{n}}$ be consistent with any experiment that only probes $\widetilde{n}$ highly entangled qubits with high fidelity, e.g. the logical qubits in a quantum computer.
This would be desirable because only $\widetilde{n} \sim \log_2 S$ highly entangled qubits would be needed to experimentally test such a model of EmQM,
  which would be experimentally relevant in the near-term if e.g. $S \sim 2^{1000}$.

Various challenges had to be overcome while constructing our model of EmQM.
In \secref{sec:ingredients}, we recount these challenges as guiding principles
  that intuitively motivate necessary ingredients for EmQM.
In \secref{sec:model}, we promote this intuition to an explicit model.
In \secref{sec:deviations}, we estimate how much our EmQM model deviates from quantum mechanics, and we numerically validate these estimates in \secref{sec:simulation}.
In \secref{sec:experiment}, we discuss possible experimental tests of EmQM models similar to the model we study.
In \secref{sec:outlook}, we mention future directions, such as how our EmQM might be modified to possibly alleviate the $S \gg 2^n$ requirement.

\section{Ingredients for \EmQM}
\label{sec:ingredients}

\subsection{Bell Nonlocality from Fast Variables}
\label{sec:fast}

Bell inequality \cite{BellTheorem,tsirelsonBound} experiments have shown that the outcomes of spacelike separated quantum measurements
  are incompatible with local hidden variable theories
  unless information can travel faster than light.
Therefore, we posit the existence of hidden \emph{fast} degrees of freedom
  that move much faster than the speed of light
  and change much more rapidly than the wavefunction.
Although we make no assumptions regarding local realism (another assumption used to derive Bell inequalities),
  we will be led to an EmQM model without local realism.
For simplicity, we consider a wavefunction for qubits.
Therefore, it is natural to take the fast variables to be classical bits.

We note that in order for a theory with faster-than-light degrees of freedom to be consistent with previous tests of Lorentz invariance \cite{LiberatiTestingLI},
  we likely also need to posit that the observed Lorentz invariance in our universe is emergent (rather than exact).
See \appref{app:Lorentz} for further discussion regarding the feasibility of this possibility.

\subsection{Linearity from Perturbative Expansion}
\label{sec:linear}

Another notable feature of Schr\"{o}dinger's equation is that it is linear in the wavefunction,
  while classical systems generically exhibit nonlinear behavior.
However, linearity is a generic result of leading-order perturbative expansions.
For example, the linear harmonic oscillator describes small oscillations of a pendulum.
The gravitational force in Newtonian gravity is a linear superposition of forces,
  which can be derived from general relativity in a certain limit where the gravitational force is weak.
Even the training dynamics of wide neural networks (for which there are many neurons per layer) can be reduced to a linear equation after a perturbative expansion about small deviations from the initial conditions \cite{NTK,wideNeuralNetworks},
  which was a significant inspiration for the EmQM model that we introduce.

We therefore posit that our model contains degrees of freedom that change so slowly with time that their dynamics can be treated in a linear approximation. If we attempt to follow the evolution for a very long time, terms nonlinear in the wavefunction may become significant, resulting in non-linear corrections \cite{WeinbergNonlinear} to the emergent Schr\"{o}dinger equation. These higher-order corrections might depend on details of the underlying classical dynamics, rather than being expressible in terms of the emergent wavefunction alone.

\subsection{Wavefunction from Probability Vector}
\label{sec:P}

An EmQM model should also explain how the quantum wavefunction is related to the classical model.
Specifying the wavefunction for $n$ qubits requires $2^n$ numbers.
This feature is reminiscent of classical probability distributions,
  which also require $2^n$ numbers for $n$ bits.
Therefore, we posit that the wavefunction is mathematically related to a probability vector $\bm{P}$.

In order to incorporate the previous two guiding principles,
  we further posit that $\bm{P}$ is a probability distribution for the fast degrees of freedom
  (i.e. the classical bits)
  and that $\bm{P}$ is determined by the slow degrees of freedom.
We further assume that the wavefunction $\Psi$ for $n$ qubits
  describes perturbations from a uniform probability vector for $n$ classical bits:
\begin{equation}
  \bm{P} = \frac{\bm{1}}{N} + \epsilon_\Psi \Psi \label{eq:P}
\end{equation}
  where $\epsilon_\Psi$ is perturbatively small.
$\bm{P}$ is a vector of probabilities for the fast degrees of freedom.
$\bm{1}$ is a vector of $N=2^n$ ones so that
  $\frac{\bm{1}}{N}$ is a uniform probability vector for $n$ bits.

We will be content to describe the evolution of the emergent wavefunction in our EmQM model, and will not discuss measurement as a separate phenomenon.
To accommodate measurements, one could adopt the Everett interpretation \cite{Everett} by including the observer and measurement apparatus as part of the physical system described by the wavefunction.
See \appref{app:measurement} for more details.

\subsubsection{Constraints}
\label{sec:constraints}

In order for $\bm{P}$ to be a valid probability vector,
  $\Psi$ must be a real vector with elements that sum to zero:
\begin{equation}
  \sum_i \Psi_i = 0 \label{eq:sumPsi}
\end{equation}
In order to preserve these constraints, the Hamiltonian $H$ in Schr\"odinger's equation \eqref{eq:Schro} must be an imaginary-valued and antisymmetric matrix
  with rows and columns that sum to zero:
\begin{equation}
\begin{aligned}
  \sum_i H_{ij} &= 0 \\
  \sum_j H_{ij} &= 0
\end{aligned} \label{eq:sumH0}
\end{equation}
In order to obtain a local EmQM model,
  we also require that every term of the Hamiltonian is local and satisfies \eqnref{eq:sumH0}.

In \appref{app:realQM}, we show that these constraints on $H$ and $\Psi$ do not result in any significant loss of generality.
In particular, given any Hamiltonian $H$ and wavefunction $\Psi(t)$ that satisfy Schr\"{o}dinger's equation,
  we find a linear mapping to a dual Hamiltonian $\widetilde{H}$ and wavefunction $\widetilde{\Psi}(t)$ that satisfy Schr\"{o}dinger's equation and the above constraints.
Furthermore, if $H$ is local, then $\widetilde{H}$ can also be chosen to be local.
Every term of $\widetilde{H}$ will also satisfy \eqnref{eq:sumH0}.

\subsection{Quantum Complexity from a Large Extra Dimension}

\begin{figure}[t!]
  \centering
  \includegraphics[width=\columnwidth]{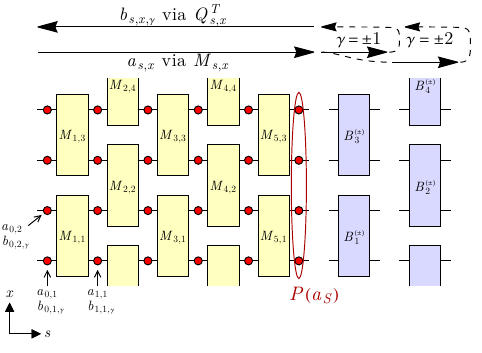} \\
  \caption{%
    A square lattice picture of our model (with $S=5$ and $n=4$),
      which exhibits EmQM on a one-dimensional boundary (circled in red) of a two-dimensional bulk.
    Classical bits $a_{s,x}$ (red dots) propagate forward through a slowly time-evolving circuit of stochastic matrices $M_{s,x}$ (yellow), and classical bits $b_{s,x,\gamma}$ propagate backwards through a time-independent circuit of permutation matrices $Q_{s,x}^T$. The emergent Schr\"{o}dinger equation describes the time evolution of the probability distribution $P(a_S)$ governing the boundary bits. The back-propagating bits determine how the stochastic matrices are updated in each time step.
    See \secref{sec:overview} for an overview.
  }\label{fig:circuit}
\end{figure}

Simulating Schr\"{o}dinger's equation with a classical computer generically requires CPU time that increases exponentially with system size.
Therefore, simulating an underlying classical EmQM model should also have a high cost.
To ensure that the EmQM model is costly to simulate,
  we posit a large extra spatial dimension of length $S$. In effect, this large dimension enables the EmQM model to describe quantum mechanics in an exponentially large Hilbert space.

In our model, the stochastic classical bits are fast in the sense that they are frequently sampled. But the emergent wavefunction is related to the probability distribution $\bm{P}$ from which these bits are sampled, where $\bm{P}$ itself evolves quite slowly in comparison. A general probability distribution on $n$ bits is parameterized by $2^n-1$ nonnegative real numbers. If we want to allow general $n$-qubit quantum pure states in our EmQM model, then many parameters must be needed to specify the stochastic process from which $\bm{P}$ arises. For this reason we assume that $\bm{P}$ is obtained by composing stochastic matrices in a very deep circuit. The number of parameters needed to parameterize the circuit is linear in its depth $S$, which we therefore assume to be exponential in $n$. We envision this circuit as extending into an auxiliary spatial dimension, not to be confused with the spatial dimensions of the emergent quantum system. 
 
Consider an EmQM model consisting of an $S+1$ by $n$ grid of bits $a_{s,x}$,
  which are depicted as red dots in \figref{fig:circuit}.
For simplicity of exposition, we focus on the EmQM of a one-dimensional chain of $n$ qubits.
Generalizations to higher spatial dimensions are straightforward.
Let $\bm{P}$ be the probability distribution for the $n$ bits at
  the $s=S$ boundary (circled in red) of the extra spatial dimension.
We then suppose that at the $s=0$ boundary,
  the $n$ bits are generated uniformly at random,
  and that the bulk dynamics interpolate between
  the uniform distribution $\frac{\bm{1}}{N}$ and $\bm{P} = \frac{\bm{1}}{N} + \epsilon_\Psi \Psi$.

Next, we take inspiration from unitary circuits,
  which can generate entangled wavefunctions from direct product states by
  repeatedly acting on pairs of qubits with unitary matrices.
Since the EmQM model involves classical bits instead of qubits,
  we instead consider a circuit of stochastic matrices $M_{s,x}$
  (yellow in \figref{fig:circuit}).
A stochastic matrix is a matrix with columns that are probability vectors,
  i.e. vectors with positive entries that sum to one.
Therefore, given two input bits,
  a $4\times4$ stochastic matrix maps those bits to a probability distribution,
  from which a new pair of bits can be sampled.
A $4\times4$ stochastic matrix can also be used to linearly map a
  probability vector for two bits to a new probability vector.
The EmQM model utilizes a circuit of stochastic matrices to sample bits from the probability vector $\bm{P}$.
We emphasize that $\bm{P}$ is not a physical degree of freedom;
  $\bm{P}$ is only implicitly defined by the circuit of stochastic matrices.\footnote{%
  The classical bits and stochastic matrices are ontic in our model.
  There are thus an infinite number of possible physical states,
      as implied by Hardy's excess baggage theorem \cite{HardyExcessBaggage}.
  However, $\bm{P}$ and the wavefunction are both fully determined by the stochastic matrices,
    which implies that our model is $\psi$-ontic \cite{LeiferOntology} in the sense that distinct wavefunctions always correspond to distinct physical states
    (of classical bits and stochastic matrices).}

The stochastic matrices $M_{s,x}$ vary slowly with the time step $\tau$, and can be decomposed as
\begin{equation}
  M_{s,x}(\tau) = Q_{s,x} + m_{s,x}(\tau), \label{eq:MQm0}
\end{equation}
  where $Q_{s,x}$ is time-independent and
  $m_{s,x}(\tau)$ is a time-dependent perturbation.
We take the $Q_{s,x}$ to be permutation matrices,
  which have the useful property that their inverse is also a stochastic matrix.\footnote{%
    More generic choices for $Q_{s,x}$ could be a useful direction for future work, which we briefly discuss in \secref{sec:modifications}.}
We will assume that at all times, the perturbation $m_{s,x}(\tau)$ is sufficiently small that we can accurately account for the evolution of $\bm{P}$ by expanding to linear order. As a result, the time evolution of the emergent wavefunction will also be described by a linear equation.

\subsection{Unitarity from Destructive Interference}
\label{sec:interference}

A final challenge is to obtain dynamics for $\bm{P}$ such that $\Psi$ in \eqnref{eq:P} undergoes a unitary evolution
  described by Schr\"{o}dinger's equation.\footnote{%
    Unitary dynamics implies many other useful properties.
    For example, Tsirelson's bound is an upper bound for how much quantum theory can violate Bell's inequality. \cite{tsirelsonBound}
    Tsirelson's bound is saturated by Schr\"odinger's equation, but Tsirelson's bound is exceeded by some alternatives of quantum theory. \cite{exceedTsirelsonBound}
    If $\Psi$ in our model obeys a unitary evolution corresponding to a local Hamiltonian, then in addition to violating Bell's inequality,
      our model would saturate Tsirelson's bound
      (in agreement with quantum theory).}
One route to realizing unitarity is to suppose that after the forward-propagating bits $a_{s,x}$ reach the $s=S$ boundary,
  the bits are transformed by a shallow stochastic circuit $B$ that encodes
  a small unitary time evolution generated by the Hamiltonian $H$.
The resulting bits could then back-propagate through the circuit (via $Q_{s,x}^T$)
  while dictating how the stochastic matrices slowly evolve such that $\bm{P}$
  undergoes the desired dynamics.

But how could a stochastic circuit encode a time evolution by a generic imaginary-valued Hamiltonian satisfying \eqnref{eq:sumH0}?
One possibility is that a pair of stochastic circuits $B^{(\pm)}$
  outputs two sets of bits with probability vectors $\bm{P}_\pm$
  that ``destructively interfere'' with each other:
\begin{equation}
  \bm{P}_+ - \bm{P}_- = (- i H  \delta_\text{t}) \cdot \bm{P}
\end{equation}
This could be achieved by stochastic circuits $B^{(\pm)}$ defined such that
\begin{equation}
  B^{(+)} - B^{(-)} = -  iH \delta_\text{t} \label{eq:Bpm}
\end{equation}
Such a decomposition is always possible for sufficiently small $\delta_\text{t}$.
For example, if
\begin{equation}
  -iH = \begin{pmatrix}
           0 & -1 & +1 & 0 \\
          +1 & 0  & -1 & 0 \\
          -1 & +1 &  0 & 0 \\
           0 &  0 &  0 & 0
        \end{pmatrix}
\end{equation}
  for $n=2$ qubits, then we can choose
\begin{equation}
\begin{aligned}
  B^{(+)} &= \begin{pmatrix}
          1-\delta_\text{t} & 0 & \delta_\text{t} & 0 \\
          \delta_\text{t} & 1-\delta_\text{t} & 0 & 0 \\
          0 & \delta_\text{t} & 1-\delta_\text{t} & 0 \\
          0 & 0 & 0 & 1
        \end{pmatrix} \\
  B^{(-)} &= \begin{pmatrix}
          1-\delta_\text{t} & \delta_\text{t} & 0 & 0 \\
          0 & 1-\delta_\text{t} & \delta_\text{t} & 0 \\
          \delta_\text{t} & 0 & 1-\delta_\text{t} & 0 \\
          0 & 0 & 0 & 1
        \end{pmatrix}
\end{aligned} \label{eq:Bexample}
\end{equation}

To illustrate how this works in a simple setting, let us continue this $n=2$ example and further suppose that $S=1$
  such that there is only a single stochastic matrix $M$, as depicted below:
\begin{equation}
  \vcenter{\hbox{\includegraphics{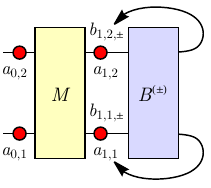}}}
\end{equation}
For each discrete time step, all of the bits $a_{s,x}$ and $b_{s,x,\pm}$ are updated.
The two input bits $a_{0,1}$ and $a_{0,2}$ are chosen uniformly at random.
The boundary bits at $s=1$ are randomly chosen from the conditional probability distributions $p(a_1|a_0) = M(a_1,a_0)$ and $p(b_{\pm}|a_1) = B^{(\pm)}(b_{\pm},a_1)$,
  where (for now) $a_s$ and $b_{\pm}$ denote pairs of bits $(a_{s,1},a_{s,2})$ and $(b_{1,1,\pm},b_{1,2,\pm})$,
  which are used to index the $4\times4$ matrices $M$ and $B^{(\pm)}$.

In order to get the desired time evolution for $\bm{P}$,
  for each discrete time step,
  $m$ in $M = Q + m$ is updated according to
\begin{equation}
  m' = m + \Delta_\text{m} (\bm{\hat{b}}_+ - \bm{\hat{b}}_-) \otimes \bm{\hat{e}} \label{eq:m'}
\end{equation}
  where $\Delta_\text{m}$ is a small constant.
$\bm{\hat{b}}_+$ denotes the basis 4-vector indexed by the two bits
  $b_{1,1,+}$ and $b_{1,2,+}$,
  and similar for $\bm{\hat{b}}_-$.
For example, $\bm{\hat{b}}_+ = (0,1,0,0)$ if $b_{1,1,+} = 0$ and $b_{1,2,+} = 1$.
$\bm{\hat{e}}$ is chosen uniformly at random from the four basis 4-vectors [$(1,0,0,0)$, $(0,1,0,0)$, etc.]
  with the constraint that $M = Q + m$ remains non-negative.
Such a choice always exists as long as the elements of $m$ (which are assumed to be small) remain smaller than 1.
The choice of $Q$ does not play an important role in this $S=1$ example.

$\bm{P}$ is the probability distribution that governs the sampled boundary bits $a_{1,1}$ and $a_{1,2}$. In addition, $\bm{P}$ itself has statistical fluctuations, because in each time step, the stochastic matrix $M$ is updated according to \eqnref{eq:m'}, where $\bm{\hat{b}}_+$ and $\bm{\hat{b}}_-$ are also stochastic variables. 
To see that \eqnref{eq:m'} results in an emergent Schr\"{o}dinger equation,
  we first calculate the expectation values (denoted by a bar) of the boundary bits:
\begin{align}
  \bm{P} = \overline{\bm{\hat{a}}_1} &= M \cdot \frac{\bm{1}_4}{4} \label{eq:P4} \\
  \overline{\bm{\hat{b}}_\pm} &= B^{(\pm)} \cdot \bm{P}
\end{align}
  where $\bm{1}_4/4$ denotes a uniform probability vector with 4 elements.
\eqnref{eq:Bpm} then implies that
\begin{equation}
  \overline{\bm{\hat{b}}_+ - \bm{\hat{b}}_-} = -i \delta_\text{t} H \cdot \bm{P} \label{eq:db4}
\end{equation}
Thus, the average change in $\bm{P}$ after one time step
  evolves according to a discrete Schr\"{o}dinger's equation:
\begin{equation}
\begin{aligned}
  \overline{\bm{P}' - \bm{P}} &= \left(\overline{m' - m}\right) \cdot \frac{\bm{1}_4}{4} \\
    &= \frac{\Delta_\text{m}}{4} \left( \overline{\bm{\hat{b}}_+ - \bm{\hat{b}}_-} \right) \\
    &= -i \Delta_\text{t} H \cdot \bm{P}
\end{aligned}
\end{equation}
  where $\Delta_\text{t} = \frac{1}{4} \delta_\text{t} \Delta_\text{m}$.
The above three equalities follow from \eqsref{eq:P4}, \eqref{eq:m'}, and \eqref{eq:db4}, respectively.
In the second equality,
  we see that the random choice of $\bm{\hat{e}}$ in \eqnref{eq:m'} does not matter
  because $\bm{\hat{e}}$ is multiplied by $\bm{1}_4$ in the first line.
Since $\bm{P}$ and $\Psi$ are linearly related by \eqnref{eq:P},
  $\Psi$ obeys Schr\"{o}dinger's equation on average.
Statistical fluctuations about this average are negligible in the small $\Delta_\text{m}$ limit.

To quantify statistical fluctuations,
  let $\tau$ denote the discrete time step of $\bm{P}^{(\tau)}$.
After $\tau$ steps, the elapsed time in the emergent quantum mechanics is $t \approx \Delta_\text{t} \tau$.
Statistical fluctuations of $\bm{P}^{(\tau)}$ grow as\footnote{%
    $\delta_t \tau$ is roughly the number of time steps for which
      $\bm{\hat{b}}_+ - \bm{\hat{b}}_- \neq 0$.
    Each of these time steps changes $m$ by $\Delta_\text{m}$.}
  $O(\Delta_\text{m} \sqrt{\delta_t \tau}) = O(\sqrt{\Delta_\text{m} t})$.
Therefore, statistical fluctuations become arbitrarily small as $\Delta_\text{m}$ decreases.

\subsection{More Qubits}
\label{sec:locality}

Finally, we must scale up the previous $S=1$ and $n=2$ example to large $S$ and $n$, as depicted in \figref{fig:circuit}.
We will now assume that the time-independent permutation matrices $Q_{s,x}$ in \eqnref{eq:MQm0} are chosen uniformly at random and independently for each $s$ and $x$. 
These random permutation matrices play an important role as they randomize the subspace in which each $m_{s,x}$ affects $\bm{P}$.
Therefore when $S n \gg N = 2^n$, the span of these subspaces covers all $N$ dimensions of $\bm{P}$.

When $S>1$, the bits $b_{S,x,\gamma}$ at the boundary will have to back-propagate through the circuit before affecting the time evolution of $M_{s,x}$.
In order to back-propagate the maximal amount of information from the boundary to the bulk,
  it would be ideal to use the inverse of the stochastic matrices $M_{s,x}^{-1}$.
But this is not possible since these inverses are generically not stochastic matrices.
However, since $m_{s,x}$ is assumed to be small,
  $M_{s,x}^{-1} \approx Q_{s,x}^T$ and therefore $Q_{s,x}^T$ can be used to deterministically back-propagate the bits.\footnote{%
    $M_{s,x}^T$ can not be used for back-propagation since
      $M_{s,x}^T$ is not guaranteed to be a stochastic matrix.
    We can not simply require $M_{s,x}^T$ to be a stochastic matrix in our model
      (which would imply that $M_{s,x}$ is doubly stochastic)
      since we require that $M_{s,x}$ maps a uniform probability distribution to a different distribution.
    But doubly stochastic matrices always map the uniform distribution to a uniform distribution
      since the rows of a doubly stochastic matrix must sum to one (by definition).}

Finally, we must split $B^{(\pm)}$ into local pieces,
  as depicted in \figref{fig:circuit}.
To be concrete, we assume that the Hamiltonian $H=\sum_x H_x$ is local such that each $H_x$ only acts on two neighboring qubits.
Generalizing \eqnref{eq:Bpm},
  $B_x^{(\pm)}$ are chosen such that
\begin{equation}
  B_x^{(+)} - B_x^{(-)} = -  iH_x \delta_\text{t} \label{eq:Bpm2}
\end{equation}
  with the constraint that the stochastic matrix
\begin{equation}
  B_x^{(\pm)} = \mathbbl{1}_4 + O(\delta_\text{t}) \label{eq:Bx0}
\end{equation}
  is the $4\times4$ identity matrix $\mathbbl{1}_4$ up to order $\delta_\text{t}$ corrections.
We then define four different depth-1 stochastic circuits:
\begin{equation}
\begin{aligned}
  B^{(\pm1)} &= \Otimes_{\text{odd  }x} B_x^{(\pm)} \\
  B^{(\pm2)} &= \Otimes_{\text{even }x} B_x^{(\pm)}
\end{aligned} \label{eq:B}
\end{equation}
  which are now guaranteed to satisfy a generalization of \eqnref{eq:Bpm}:
\begin{equation}
  +B^{(+1)} - B^{(-1)} + B^{(+2)} - B^{(-2)} = -i H \delta_\text{t} + O(\delta_\text{t}^2) \label{eq:Bsum}
\end{equation}
Since there are now four different flavors of $B^{(\gamma)}$,
  we also take four different flavors of back-propagating bits $b_{s,x,\gamma}$ with $\gamma=\pm1$ and $\gamma=\pm2$.
With these ingredients combined,
  we arrive at a local model of emergent quantum mechanics.

\section{\EmQM\ Model}
\label{sec:model}

\subsection{Overview}
\label{sec:overview}

In summary, we introduce a two-dimensional local classical model for which a wavefunction $\Psi$ of qubits is encoded in a probability distribution $\bm{P} = \frac{\bm{1}}{N} + \epsilon_\Psi \Psi$ [\eqnref{eq:P}]
  for $n$ classical bits on a one-dimensional boundary of the model.
Generalizing to higher dimensions or $Z_k$ qudits (rather than $Z_2$ qubits) is straightforward.
For each discrete time step,
  a bit string $a_S$ (e.g. $a_S = 011010$) is generated with probability $P(a_S)$ at the $s=S$ boundary.
Our model exhibits EmQM in the sense that $\bm{P}$
  [i.e. $P(a_S)$ viewed as a vector with $N = 2^n$ components]
  evolves according to
\begin{equation}
  \partial_t \bm{P} \approx -iH \cdot \bm{P} \label{eq:SchroP}
\end{equation}
  where $H$ is imaginary-valued and obeys \eqnref{eq:sumH0}.
Schr\"{o}dinger's equation \eqref{eq:Schro} for $\Psi$
  then follows from $\bm{P} = \frac{\bm{1}}{N} + \epsilon_\Psi \Psi$ [\eqnref{eq:P}].
We will derive \eqnref{eq:SchroP} from a well-controlled perturbation theory,
  which we verify numerically.
Furthermore, the approximate equality in \eqnref{eq:SchroP} becomes exact in a well-defined limit.

\begin{table*}[p]
\setlength{\tabcolsep}{6pt}
\renewcommand{\arraystretch}{1.1}
\begin{tabular}{c|l|l}
variable & definition & reference \\ \hline
$n$ & number of bits output by circuit at each time step & \secref{sec:overview} \\
$N = 2^n$ & number of bit strings for $n$ bits; $x=1,2,\ldots,n$ & \secref{sec:overview} \\
$S$ & length of extra dimension; $s=1,2,\ldots,S$ & \secref{sec:overview} \\
$\delta_\text{t}$ & raw time step (not to be confused with $\Delta_\text{t}$) & \secref{sec:interference} \\
$m_0$ & approximate Hilbert–Schmidt norm of each $m_{s,x}$ & \eqnref{eq:p} \\
$\Delta_\text{m}$ & step size for $m_{s,x}$ for each time step & \eqnref{eq:dm} \\
$\epsilon_0$ & small parameter used to parameterize the above four constants & \eqnref{eq:epsilon} \\
$H$, $H_x$ & effective local Hamiltonian $H = \sum_x H_x$ & \secref{sec:overview} \\
$G$, $G_x$ & real-valued: $G=-iH$, $G_x=-iH_x$ & \eqnref{eq:G} \\
\hline
$\Delta_\text{t}$ & effective EmQM time step: $t \approx \Delta_t \tau$ & \eqnref{eq:Delta_t} \\
$\epsilon_\Psi$ & small constant relating $\bm{P} - \frac{\bm{1}}{N} \approx \epsilon_\Psi \Psi$ & \eqsref{eq:P} and \eqref{eq:epsPsi} \\
$B_x^{(\pm)}$ & $4\times4$ stochastic matrix used to sample $b_{s,x,\pm\gamma}$ & \eqsref{eq:Bpm2}, \eqref{eq:bS'}, and \eqref{eq:Bx} \\
$B^{(\gamma)}$ & depth-1 stochastic circuit; related to Hamiltonian & \eqsref{eq:B} and \eqref{eq:Bsum} \\
$\Delta_\text{P}$ & rough 2-norm of $\overline{\bm{P}^{(\tau)} - \bm{P}^{(\tau-1)}}$ & \eqnref{eq:DeltaP} \\
$P(a_S)$ & probability the circuit outputs bit string $a_S$ & \secref{sec:overview} \\
$\bm{P} = \bm{P}_S$ & expectation value of $\bm{\hat{a}}_S$, or $P(a_S)$ viewed as a probability vector & \eqsref{eq:Ps}, and \eqref{eq:p} \\
$\varepsilon(t)$ & EmQM deviation $||\Psi(t) - \Psi_\text{QM}(t)||$ from quantum mechanics (QM) & \eqsref{eq:epst} and \eqref{eq:epsLag} \\
$\varepsilon_\text{m}(t)$, $\varepsilon_\text{t}(t)$ & deviations due to finite $m_0$ and $\delta_\text{t}$ & \eqsref{eq:eps m} and \eqref{eq:eps t} \\
$\varepsilon_\text{S}(t)$, $\varepsilon_\text{stat}(t)$ & deviations due to finite $S$ and statistical fluctuations & \eqsref{eq:eps S} and \eqref{eq:stat} \\
\hline
$\tau$ & integer-valued time step & \secref{sec:circuit} \\
$a_{s,x}$ & a forward-propagating stochastic bit & \secref{sec:overview} \\
$a_{s,\xx}$ & pair of bits $(a_{s,x}, a_{s,x+1})$ & above \eqnref{eq:Ma} \\
$a_s$ & bit string ($a_{s,1}, a_{s,2}, \ldots, a_{s,n}$) & below \eqnref{eq:uniform} \\
$\bm{\hat{a}}_s$ & length-$n$ basis vector
  indexed by the bit string $a_s$ & below \eqnref{eq:db} \\
$b_{s,x,\gamma}$ & a backward-propagating stochastic bit & \secref{sec:overview} \\
$b_{s,\xx,\gamma}$ & a pair of bits $(b_{s,x,\gamma}, a_{s,x+1,\gamma})$ & above \eqnref{eq:bS'} \\
$\bm{\hat{b}}_{s,\xx,\gamma}$ & $(1,0,0,0)$, $(0,1,0,0)$, $(0,0,1,0)$, or $(0,0,0,1)$ when $b_{s,\xx,\gamma}=00$, $01$, $10$, or $11$ & below \eqnref{eq:dm2} \\
$b_{s,\gamma}$ & bit string $(b_{s,1,\gamma},b_{s,2,\gamma},\ldots,b_{s,n,\gamma})$ & below \eqnref{eq:dm2} \\
$\bm{\hat{b}}_{s,\gamma}$ & length-$n$ basis vector
  indexed by the bit string
  $b_{s,\gamma}$ & below \eqnref{eq:dm2} \\
$M_{s,x}$ & $4\times4$ stochastic matrices, which define the classical circuit & \secref{sec:overview} \\
$M_s$ &Kronecker product of $M_{s,x}$ with fixed $s$ & \eqnref{eq:Ms} \\
$M_{S \leftarrow s}$ & $= M_S \cdot M_{S-1} \cdots M_{s+1}$ & \eqnref{eq:MSs} \\
$m_{s,x}$ & perturbations to $M_{s,x} = Q_{s,x} + m_{s,x}$ & \eqnref{eq:MQm} \\
\hline
$Q_{s,x}$ & time-independent $4\times4$ permutation matrices & below \eqnref{eq:MQm} \\
$Q_s$ & Kronecker product of $Q_{s,x}$ with fixed $s$ & \eqnref{eq:Qs} \\
$Q_{S\leftarrow s}$ & $= Q_S \cdot Q_{S-1} \cdots Q_{s+1}$ & \eqnref{eq:QSs} \\
\hline
$\bm{1}$ or $\bm{1}_n$ & vector of $n$ ones & \\
$\mathbbl{1}$ or $\mathbbl{1}_n$ & $n\times n$ identity matrix & \\
$\mathcal{P}_x$ & projects out all bits except $x$ and $x+1$ & \eqnref{eq:Px} \\
$\mathcal{W}$ & sum of conjugated projectors $\mathcal{P}_x$ & \eqsref{eq:SchroW} and \eqref{eq:W} \\
\end{tabular}
\caption{A table of notation used throughout the main text [with the exception of \secref{sec:interference} which uses some simplified notations].
Notations are grouped as follows: model parameters, dependent variables, dynamical variables, randomly-initialized constants, and other constants.} \label{tab:notation}
\end{table*}

\figref{fig:circuit} summarizes the basic structure of our EmQM model.
The model consists of a square lattice of classical bits
  $a_{s,x}$ and $b_{s,x,\gamma}$ (red dots)
  that propagate through a brick circuit of slowly-varying stochastic matrices $M_{s,x}$ (yellow).
For each discrete time step, pairs of classical forward-propagating bits ($a_{s,x}$ and $a_{s,x+1}$)
  are randomly sampled \eqref{eq:Ma} from probability distributions conditioned on the pair of bits to the left ($a_{s-1,x}$ and $a_{s-1,x-1}$).
The conditional probabilities are encoded in the stochastic matrices $M_{s,x}$,
  which are perturbatively close to time-independent permutation matrices $Q_{s,x} \approx M_{s,x}$.
Also at each time step, pairs of backward-propagating bits ($b_{s,x,\gamma}$ and $b_{s,x+1,\gamma}$)
  are deterministically replaced \eqref{eq:Qb} by the permutation $Q_{s,x}^T$
  of bits to their right ($b_{s+1,x}$ and $b_{s+1,x+1}$).
The bits $a_{0,x}$ at the left $s=0$ boundary are uniformly initialized at random.
The bits $b_{S,x,\gamma}$ at the right $s=S$ boundary
  result from applying \eqref{eq:bS'} a shallow stochastic circuit \eqref{eq:B}
  to the bits $a_{S,x}$ at $s=S$.
The shallow circuit is related \eqref{eq:Bsum} to the Hamiltonian
  and consists of a layer of time-independent stochastic matrices $B_x^{(\pm)}$ (blue).
The bits $a_{S,x}$ on the $s=S$ boundary (circled in red) follow a probability distribution $\bm{P}$ defined by a tensor network product \eqref{eq:Ps} of the matrices $M_{s,x}$.
The time-evolution of $M_{s,x}$ depends \eqref{eq:dm} on back-propagating bits $b_{s,x,\gamma}$,
  which effectively back-propagate the result of a small Hamiltonian time evolution on $\bm{P}$,
  such that $\bm{P}$ and $\Psi$ approximately obey Schr\"{o}dinger's equation.
See \tabref{tab:notation} for a notational reference.

\subsection{Stochastic Circuit}
\label{sec:circuit}

We now define and study the EmQM model in greater detail.
Each lattice site hosts a classical bit $a_{s,x}=0,1$,
  where $s = 0,1,\ldots,S$, and $x = 1,\ldots,n$ index the different lattice sites.
$s$ is the coordinate for an extra dimension,
  while $x$ is the coordinate for the spatial dimension on which the emergent qubits live.
Our model evolves in discrete time steps indexed by integer-valued $\tau$, while
  $t$ is reserved for the emergent time variable of the emergent quantum mechanics.
$a_{s,x}^{(\tau)}$ denotes the value of the bit $a_{s,x}$ at time step $\tau$,
  and similar for other time-dependent variables.
In contexts where the time step is not important,
  we often omit the $(\tau)$ superscript to avoid clutter.

To avoid clutter, we denote a pair of bits $(a_{s,x}, a_{s,x+1})$ as $a_{s,\xx}$,
  where $\xx$ is shorthand for $(x,x+1)$.
For each time step $\tau$, the classical bits $a_{s,x}^{(\tau-1)}$ with $1 \leq s \leq S$
  are stochastically updated to $a_{s,x}^{(\tau)}$ with conditional probabilities
  that are conditioned on the bits $a_{s-1,\xx}^{(\tau-1)}$.
These conditional probabilities are given by time-dependent stochastic matrices $M_{s,x}$:
\begin{gather}
  p(a_{s,\xx}^{(\tau)} | a_{s-1,\xx}^{(\tau-1)}) =
    M_{s,x}^{(\tau-1)}(a_{s,\xx}^{(\tau)}, a_{s-1,\xx}^{(\tau-1)})
 \label{eq:Ma}\\
  \vcenter{\hbox{\includegraphics{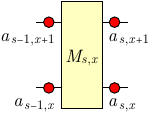}}} \nonumber
\end{gather}
For each $(s,x)$ with $s=1,2,\ldots,S$ and $s-x$ even,
  $M_{s,x}$ is a $4\times4$ stochastic matrix.
That is, the columns of $M_{s,x}$ are probability vectors;
  i.e. $M_{s,x}$ has positive elements and columns that sum to 1.
We index the $4\times4$ matrices using pairs of bits,
  such that for fixed $a_{s-1,\xx}^{(\tau-1)}$,
  $M_{s,x}^{(\tau-1)}(a_{s,\xx}^{(\tau)}, a_{s-1,\xx}^{(\tau-1)})$ is a probability distribution for $a_{s,\xx}^{(\tau)}$.
The bits along the $s=0$ line at the beginning of the circuit are randomly sampled with equal probability:
\begin{equation}
  p(a_{0,x}) = 1/2 \label{eq:uniform}
\end{equation}

Let $a_s \equiv (a_{s,1}, a_{s,2}, \ldots, a_{s,n})$ denote the string of bits along a column of fixed $s$.
For each $s$, let $\bm{P}_s$ denote the vector of probabilities for the different bit strings $a_s$.
Then $\bm{P}_s^{(\tau)}$ can be expressed recursively as
\begin{equation}
\begin{aligned}
  \bm{P}_s^{(\tau)} &= M_s^{(\tau-1)} \cdot \bm{P}_{s-1}^{(\tau-1)} \\
  \bm{P}_0^{(\tau)} &= \frac{\bm{1}}{N}
\end{aligned} \label{eq:Ps}
\end{equation}
  where $M_s$ denotes the Kronecker product of stochastic matrices with fixed $s$:
\begin{equation}
  M_s^{(\tau)} = \Otimes_x^{\substack{\text{even}\\s-x}} M_{s,x}^{(\tau)}. \label{eq:Ms}
\end{equation}
The notation on the right-hand side means we take the Kronecker product of all $M_{s,x}^{(\tau)}$ for even $x$ when $s$ is even 
  (or for odd $x$ when $s$ is odd), producing the brickwork circuit shown in \figref{fig:circuit}.
The bit string probabilities at the end of the circuit are given by $\bm{P} = \bm{P}_S$ (circled in red in \figref{fig:circuit}).

\subsection{Perturbative Expansion}

To gain analytical tractability,
  we assume that the stochastic matrices are very close to permutation matrices:
\begin{equation}
  M_{s,x}^{(\tau)} = Q_{s,x} + m_{s,x}^{(\tau)} \label{eq:MQm}
\end{equation}
Each $Q_{s,x}$ is a randomly-chosen (for each $s$ and $x$) but time-independent $4\times4$ permutation matrix,
  while $m_{s,x}^{(\tau)}$ is a small time-dependent perturbation.
A permutation matrix is a stochastic matrix where all elements are either 0 or 1.
A matrix is a permutation matrix if and only if it is both stochastic and orthogonal.
The dynamics of $m_{s,x}$ will be constrained such that $M_{s,x}$ remains a stochastic matrix (with non-negative components).

Since they are orthogonal, the permutation matrices have the effect of a basis transformation.
Similar to \eqnref{eq:Ms},
  let $Q_s$ denote the Kronecker product of permutation matrices with fixed $s$:
\begin{equation}
  Q_s = \Otimes_x^{\substack{\text{even}\\s-x}} Q_{s,x} \label{eq:Qs}
\end{equation}
Let $Q_{S \leftarrow s}$ denote the product
\begin{equation}
  Q_{S \leftarrow s} = Q_S \cdot Q_{S-1} \cdots Q_{s+1} \label{eq:QSs}
\end{equation}
  such that it encodes the change of basis from the column of bits at $s$ to the end at $s=S$.

Expanding $\bm{P}$ to first order in the perturbations $m_{s,x}$ results in:
\begin{gather}
  \bm{P}^{(\tau)} = \frac{\bm{1}}{N} + \sum_{s=1}^S \sum_x^{\substack{\text{even}\\s-x}}
    Q_{S \leftarrow s} \cdot m_{s,x}^{(\tau_s)} \cdot \frac{\bm{1}}{N} + O(m_0^2) \label{eq:p}\\
  \vcenter{\hbox{\includegraphics{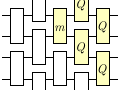}}} \nonumber
\end{gather}
  where $\tau_s = \tau-1-(S-s)$ results from a time delay, and
  $O(m_0^2)$ denotes terms that are quadratic in $m_{s,x}$.
For simplicity, we assume that each $m_{s,x}$ has Hilbert–Schmidt norm roughly equal to $m_0$.
In \secref{sec:deviations},
  we will address more precisely the effects of the non-linear $O(m_0^2)$ terms
  and other approximations that we make.

As elucidated in the above picture, $m_{s,x}^{(\tau_s)}$ affects the probabilities for two bits at $s$ and $(x,x+1)$;
  these probabilities span a 4-dimensional subspace of the $N$-dimensional vector space of probabilities $\bm{P}_s$ (after averaging over the uniform distribution of input bits).
This subspace is then scrambled into a different basis by
  the product $Q_{S \leftarrow s}$ of many random (but time-independent) permutation matrices.
If $n S \gg N = 2^n$, then the linear combination of these subspaces will span the entire $N$-dimensional vector space.
Thus, the sum of contributions from all $m_{s,x}$ can encode any wavefunction $\Psi$ [that is real-valued and satisfies \eqnref{eq:sumPsi}],
  where $\bm{P} = \frac{\bm{1}}{N} + \epsilon_\Psi \Psi$ [\eqnref{eq:P}].

\subsection{Time Evolution}

We now want to give the perturbations $m_{s,x}$ a time evolution that leads to an emergent Schr\"{o}dinger equation;
  i.e. $\bm{P}^{(\tau)} - \bm{P}^{(\tau-1)} \propto -iH \cdot \bm{P}$.
To accomplish this,
  we introduce a set of backward-propagating bits $b_{s,x,\gamma}$ with $\gamma=\pm1, \pm2$.
At the $s=S$ boundary, the pair of bits $b_{S,\xx,\pm\gamma}^{(\tau)}$ are
  randomly sampled with probabilities that are conditioned on the bits $a_{S,\xx}^{(\tau-1)}$.
These conditional probabilities are encoded in
  time-independent $4\times4$ stochastic matrices
  $B_x^{(\pm)}$ (blue in \figref{fig:circuit}):
\begin{equation}
  p(b_{S,\xx,\pm\gamma}^{(\tau)}| a_{S,\xx}^{(\tau-1)}) = B_x^{(\pm)}(b_{S,\xx,\pm\gamma}^{(\tau)}, a_{S,\xx}^{(\tau-1)}) \label{eq:bS'}
\end{equation}
  where $\gamma=1$ if $x$ is odd, else $\gamma=2$.

The bits $b_{s,\xx,\gamma}$ deterministically propagate backwards through the circuit
  via the transposed permutation matrices $Q_{s,x}^T$:
\begin{equation}
  \bm{\hat{b}}_{s,\xx,\gamma}^{(\tau)} = Q_{s,x}^T \cdot \bm{\hat{b}}_{s+1,\xx,\gamma}^{(\tau-1)} \label{eq:Qb}
\end{equation}
The hat over $\bm{\hat{b}}_{s,\xx,\gamma}^{(\tau)}$ denotes the basis 4-vector indexed by the two bits $b_{s,\xx,\gamma}^{(\tau)}$;
  i.e. $\bm{\hat{b}}_{s,\xx,\gamma}^{(\tau)} = (1,0,0,0)$ if $b_{s,\xx,\gamma}^{(\tau)} = 00$, and
       $\bm{\hat{b}}_{s,\xx,\gamma}^{(\tau)} = (0,1,0,0)$ if $b_{s,\xx,\gamma}^{(\tau)} = 01$, etc.

The time-evolution of the perturbations $m_{s,x}$ is stochastic and
  depends on the back-propagating bits as follows [extending \eqnref{eq:m'}]:
\begin{equation}
  m_{s,x}^{(\tau)} = m_{s,x}^{(\tau-1)} + \Delta_\text{m}
    \sum_{\gamma=1,2} \left( \bm{\hat{b}}_{s,\xx,+\gamma}^{(\tau)} - \bm{\hat{b}}_{s,\xx,-\gamma}^{(\tau)} \right)
    \otimes \bm{\hat{e}}_{s-1,x,\gamma}^{(\tau)} \label{eq:dm}
\end{equation}
  where $\Delta_\text{m}$ is a small positive constant.
Similar to $\bm{\hat{b}}_{s,\xx,\gamma}^{(\tau)}$ (defined in the previous paragraph),
  $\bm{\hat{e}}_{s-1,x,\gamma}^{(\tau)}$ is also one of the four basis 4-vectors,
  except it is chosen uniformly at random from the set of basis 4-vectors
  that keep $M_{s,x}^{(\tau)} = Q_{s,x} + m_{s,x}^{(\tau)}$ non-negative.
Such a choice always exists as long as the elements of $m_{s,x}^{(\tau)}$ (which are assumed to be small) remain smaller than 1.
The random choice of $\bm{\hat{e}}_{s-1,x,\gamma}^{(\tau)}$ has little effect
  since $m_{s,x}^{(\tau)}$ enters \eqnref{eq:p} after
  right-multiplication by $\bm{1}$.

The following quantity will play an important role:
\begin{equation}
  \left(m_{s,x}^{(\tau)} - m_{s,x}^{(\tau-1)} \right) \cdot \frac{\bm{1}_4}{4} =
    \frac{\Delta_\text{m}}{4} \sum_{\gamma = 1,2} \left( \bm{\hat{b}}_{s,\xx,+\gamma}^{(\tau)} - \bm{\hat{b}}_{s,\xx,-\gamma}^{(\tau)} \right) \label{eq:dm1}
\end{equation}
  where $\frac{\bm{1}_4}{4}$ is a length-4 uniform probability vector.
When $m_{s,x}^{(\tau)} - m_{s,x}^{(\tau-1)}$ acts on two bits of a
  uniform probability vector of length $N=2^n$, the result is similar:
\begin{align}
  & \left(m_{s,x}^{(\tau)} - m_{s,x}^{(\tau-1)} \right) \cdot \frac{\bm{1}_N}{N} \nonumber\\
    &\quad = \frac{\bm{1}_{2^{x-1}}}{2^{x-1}} \otimes \left[ \left(m_{s,x}^{(\tau)} - m_{s,x}^{(\tau-1)} \right) \cdot \frac{\bm{1}_4}{4} \right] \otimes \frac{\bm{1}_{2^{n-x-1}}}{2^{n-x-1}} \\
    &\quad = \frac{\Delta_\text{m}}{4} \sum_{\gamma = 1,2} \mathcal{P}_x \cdot \left( \bm{\hat{b}}_{s,+\gamma}^{(\tau)} - \bm{\hat{b}}_{s,-\gamma}^{(\tau)} \right) \label{eq:dm2}
\end{align}
In the last line,
  $\bm{\hat{b}}_{s,\gamma}$ denotes a length-$n$ basis vector
  indexed by the bit string
  $b_{s,\gamma} = (b_{s,1,\gamma},b_{s,2,\gamma},\ldots,b_{s,n,\gamma})$.
For example if $n=3$, then
  $\bm{\hat{b}}_{s,\gamma} = (1,0,0,0,0,0,0,0)$ if $b_{s,\gamma} = 000$, and
  $\bm{\hat{b}}_{s,\gamma} = (0,1,0,0,0,0,0,0)$ if $b_{s,\gamma} = 001$, etc.
$m_{s,x}^{(\tau)} - m_{s,x}^{(\tau-1)}$ only acts on bits $x$ and $x+1$.
Therefore the other bits are projected out by the projection matrix
\begin{equation}
  \mathcal{P}_x = \Otimes_{x'=1}^n \begin{cases}
    \quad \begin{pmatrix} 1 & 0 \\ 0 & 1   \end{pmatrix} &
      \text{\parbox[c][][t]{0.09\textwidth}{$x' = x \;$ or \\ $x' = x+1$}} \\
    \begin{pmatrix} 1/2 & 1/2 \\ 1/2 & 1/2 \end{pmatrix} & \text{otherwise}
    \end{cases} \label{eq:Px}
\end{equation}
Expressing \eqnref{eq:dm2} in this way will be useful later on.

\subsection{\texorpdfstring{$B_x^{(\pm)}$}{B} Encode the Hamiltonian}

Given a Hamiltonian $H = \sum_x H_x$,
  we now wish to choose $B_x^{(\pm)}$ such that inserting \eqnref{eq:dm2} into \eqref{eq:p} yields a discrete Schr\"{o}dinger equation:
  $\bm{P}^{(\tau)} - \bm{P}^{(\tau-1)} \propto G \cdot \bm{P}$ where we define
\begin{equation}
  G = -iH \label{eq:G}
\end{equation}
  and $G_x = -iH_x$.

We assume $H_x$ is Hermitian, imaginary-valued, and acts only on two neighboring qubits,
  which implies that $G_x$ is a real antisymmetric $4\times4$ matrix.
In accordance with \eqnref{eq:sumH0}, we further assume that $G_x$ has rows and columns that sum to zero.
We also decompose $G_x = G_x^{(+)} - G_x^{(-)}$ such that
  $G_x^{(+)}$ and $G_x^{(-)}$ only have non-negative elements.

We require that $B_x^{(\pm)}$ satisfy \eqsref{eq:Bpm2} and \eqref{eq:Bx0} so that these matrices encode the Hamiltonian, as in \eqnref{eq:Bsum}.
To achieve this for a general (geometrically two-local) Hamiltonian,
  we choose $B_x^{(\pm)}$ to be the following $4\times4$ matrix:
\begin{align}
  B_x^{(\pm)}(b_\pm,a) &= \delta_\text{t} G_x^{(\pm)}(b_\pm,a) \label{eq:Bx}\\
                   &\quad + \begin{cases} 1-\delta_\text{t} \, g_x(a) & b_\pm = a \\
                                          0 & b_\pm \neq a \end{cases} \nonumber
\end{align}
  where $g_x(a)$ is a column sum of $G_x^{(\pm)}$:
\begin{equation}
  g_x(a) = \sum_{b=00,01,10,11} \, G_x^{(\pm)}(b, a) \label{eq:g}
\end{equation}
``$b_+$'', ``$b_-$'', and ``$a$'' each denote a pair of bits,
  which respectively correspond to
  $b_{S,\xx,+\gamma}$, $b_{S,\xx,-\gamma}$, and $a_{S,\xx}$
  in \eqnref{eq:bS'}.
$B_x^{(\pm)}$ are $4\times4$ stochastic matrices,
  which we view as a probability vector for 2 bits ($b_\pm$) given two bits ($a$).
Either $G_x^{(+)}$ or $G_x^{(-)}$ can be used in the right-hand-side of \eqnref{eq:g};
  both give the same result since the column sum of $G_x = G_x^{(+)} - G_x^{(-)}$ is zero.
$B_x^{(\pm)}$ is a stochastic matrix as long as $\delta_\text{t}$ is sufficiently small.
See \eqnref{eq:Bexample} for an example.

Importantly, since this choice of $B_x^{(\pm)}$ implies \eqnref{eq:Bsum},
  the bits $b_{S,\gamma}$ encode a short time evolution by the Hamiltonian:
\begin{align}
  &\sum_{\gamma=1,2} \overline{\bm{\hat{b}}_{S,+\gamma}^{(\tau)} - \bm{\hat{b}}_{S,-\gamma}^{(\tau)}} \nonumber\\
  &\quad = (+B^{(+1)} - B^{(-1)} + B^{(+2)} - B^{(-2)}) \cdot \overline{\bm{\hat{a}}_S^{(\tau-1)}} \nonumber\\
  &\quad = (-i\delta_\text{t} H) \cdot \bm{P}^{(\tau-1)} + O(\delta_\text{t}^2) \label{eq:db}
\end{align}
The first equality follows from \eqnref{eq:bS'},
  where $B^{(\gamma)}$ was defined in \eqnref{eq:B}.
Similar to the $\bm{\hat{b}}_{s,\gamma}$ notation,
  $\bm{\hat{a}}_s$ denotes a length-$n$ basis vector
  indexed by the bit string $a_s$.
The final line follows from \eqnref{eq:Bsum}
  and the definition $\bm{P}^{(\tau)} = \overline{\bm{\hat{a}}_S^{(\tau)}}$.

\subsection{Emergent Quantum Mechanics}

Now that we have specified the EmQM model,
  we show that it can exhibit emergent quantum mechanics.
We begin by evaluating the expectation value of $\bm{P}^{(\tau)} - \bm{P}^{(\tau-1)}$
  using \eqnref{eq:p}:
\begin{align}
  &\overline{\bm{P}^{(\tau)} - \bm{P}^{(\tau-1)}} = \label{eq:dP}\\
    &\quad \sum_{s=1}^S \sum_x^{\substack{\text{even}\\s-x}}
      Q_{S \leftarrow s} \cdot \left(\overline{m_{s,x}^{(\tau_s)} - m_{s,x}^{(\tau_s-1)}}\right) \cdot \frac{\bm{1}}{N} + O(m_0^2) \nonumber
\end{align}
We can simplify the summand as follows:
\begin{align}
  &\left(\overline{m_{s,x}^{(\tau_s)} - m_{s,x}^{(\tau_s-1)}}\right) \cdot \frac{\bm{1}}{N} \nonumber\\
    &\quad = \frac{\Delta_\text{m}}{4} \sum_{\gamma=1,2} \mathcal{P}_x \cdot Q_{S \leftarrow s}^T \cdot
      \left( \overline{\bm{\hat{b}}_{S,+\gamma}^{(\tau_s')} - \bm{\hat{b}}_{S,-\gamma}^{(\tau_s')}} \right) \label{eq:dmN 1}\\
    &\quad = \frac{\Delta_\text{m}}{4} \mathcal{P}_x \cdot Q_{S \leftarrow s}^T \cdot (-i\delta_\text{t} H) \cdot \overline{\bm{P}^{(\tau_s'-1)}} + O(\delta_\text{t}^2) \label{eq:dmN 2}
\end{align}
\eqnref{eq:dmN 1} is obtained from \eqnref{eq:dm2}
  after replacing $\bm{\hat{b}}_{s,\gamma}^{(\tau_s)} = Q_{S \leftarrow s}^T \cdot \bm{\hat{b}}_{S,\gamma}^{(\tau_s')}$,
  where $\tau_s' = \tau_s - (S-s) = \tau - 1 - 2(S-s)$.
\eqnref{eq:dmN 2} follows from \eqnref{eq:db}.

We assume $\Delta_\text{m} \ll S^{-1}$ such that the slow variables $m_{s,x}$ change very little
  over $S$ time steps so that we can approximate
  $\bm{P}^{(\tau_s'-1)} = \bm{P}^{(\tau-1)} + O(S \Delta_\text{m})$.
Inserting \eqnref{eq:dmN 2} into \eqref{eq:dP} results in
\begin{equation}
\begin{aligned}
  \overline{\bm{P}^{(\tau)} - \bm{P}^{(\tau-1)}} &= \frac{\Delta_\text{m}}{4} \delta_\text{t} \, \mathcal{W} \cdot (-iH) \cdot \overline{\bm{P}^{(\tau-1)}} \\
   &\quad+ O(m_0^2,\delta_\text{t}^2,\Delta_\text{m}^2)
\end{aligned} \label{eq:dP2}
\end{equation}
  where $\mathcal{W}$ is a sum of conjugated projectors $\mathcal{P}_x$ [\eqnref{eq:Px}]:
\begin{equation}
  \mathcal{W}
    = \sum_{s=1}^S \sum_x^{\substack{\text{even}\\s-x}}
        Q_{S \leftarrow s} \cdot \mathcal{P}_x \cdot Q_{S \leftarrow s}^T \label{eq:W}
\end{equation}
The error estimates $O(m_0^2,\delta_\text{t}^2,\Delta_\text{m}^2)$ in \eqnref{eq:dP2} neglect factors of $S$ and $N$,
  which will be accounted for in \secref{sec:deviations}.

If $S n \gg N$, then $\mathcal{W}$ approaches a simple form:
\begin{equation}
    \mathcal{W} = \underbrace{\frac{S n}{2} \mathcal{P}_{\bm{1}} + \frac{S n}{2} \frac{3}{N-1} \mathcal{P}_{\bm{1}}^\perp}_{\mathcal{W}_0} + O\!\left(\sqrt{\frac{S n}{N}}\right) \label{eq:W2}
\end{equation}
  where the matrix $\mathcal{P}_{\bm{1}} = \frac{\bm{1} \otimes \bm{1}}{N}$
  projects onto the one-dimensional subspace spanned by  $\bm{1}$
  (i.e. the vector of $N$ ones)
  and $\mathcal{P}_{\bm{1}}^\perp = \mathbbl{1} - \mathcal{P}_{\bm{1}}$ projects into the orthogonal subspace.
$\mathbbl{1}$ is an $N \times N$ identity matrix.

The first term in \eqnref{eq:W2} follows from multiplying \eqnref{eq:W} by $\bm{1}$, and noting that $Sn/2$ is the number of terms summed in \eqnref{eq:W}.
To understand the following two terms,
  note that within the $(N-1)$-dimensional subspace orthogonal to $\bm{1}$,
  the set of permutation matrices is a unitary 1-design \cite{unitaryDesign}
  up to a constant factor.
That is,
\begin{equation}
  \Exp_Q \; Q_\perp \otimes Q_\perp^T =
  \frac{1}{1-N^{-1}} \Exp_U \; U_\perp \otimes U_\perp^\dagger \label{eq:design}
\end{equation}
  where $Q_\perp = \mathcal{P}_{\bm{1}}^\perp \cdot Q \cdot \mathcal{P}_{\bm{1}}^\perp$ and   $U_\perp = \mathcal{P}_{\bm{1}}^\perp \cdot U \cdot \mathcal{P}_{\bm{1}}^\perp$, and
  $\tExp_Q$ averages over all permutation matrices $Q$
  while $\tExp_U$ averages over all unitary matrices.\footnote{%
    \eqnref{eq:design} also holds if $\tExp_U$ only averages over orthogonal matrices (since the set of orthogonal matrices is also a unitary 1-design)
      or if $\Exp_Q$ only averages over permutation matrices that are affine transformations
        (which is equivalent to the set of matrices generated by permutation matrices that only act on two bits \cite{AaronsonAffine,AffineBook}).
    \eqnref{eq:design} follows after evaluating
      $\Exp_Q Q_{ij} Q_{kl} = \frac{1}{N}  \mathbbl{1}_{ik} \mathbbl{1}_{jl} + \frac{1}{N(N-1)}  (\bm{1}\otimes\bm{1}-\mathbbl{1})_{ik} (\bm{1}\otimes\bm{1}-\mathbbl{1})_{jl}$ and $\Exp_U U_{ij} \otimes U_{kl}^* = \frac{1}{N} \mathbbl{1}_{ik} \mathbbl{1}_{jl}$.
      The components $\mathbbl{1}_{ij}$ of the identity matrix $\mathbbl{1}$ are equivalent to a Kronecker delta: $\delta_{ij} = \mathbbl{1}_{ij}$.}
Our numerical experiments show that the permutation matrices $Q_{S \leftarrow s}$
  with $S-s \gg n$ summed in \eqnref{eq:W} are an approximate 1-design in the same sense \cite{PermutationsMix,BrandaoApproximateDesigns,DerandomizedPermutations}:
\begin{equation}
  \Exp_{Q_{s,x}} \; (Q_{S \leftarrow s})_\perp \otimes (Q_{S \leftarrow s})_\perp^T \approx
  \frac{1}{1-N^{-1}} \Exp_U \; U_\perp \otimes U_\perp^\dagger \label{eq:approxDesign}
\end{equation}
  where $(Q_{S \leftarrow s})_\perp = \mathcal{P}_{\bm{1}}^\perp \cdot Q_{S \leftarrow s} \cdot \mathcal{P}_{\bm{1}}^\perp$.
The above equation allows us to approximate the average of $\mathcal{W}$ (in the subspace orthogonal to $\bm{1}$) as:
\begin{align}
  \mathcal{P}_{\bm{1}}^\perp \cdot \overline{\mathcal{W}} \cdot \mathcal{P}_{\bm{1}}^\perp
  &\approx \frac{S}{2} \frac{1}{1-N^{-1}} \sum_x \Exp_U \; U_\perp \cdot \mathcal{P}_x \cdot U_\perp^\dagger \nonumber\\
  &= \frac{S}{2} \frac{1}{N-1} \sum_x \mathcal{P}_{\bm{1}}^\perp \cdot \mathcal{P}_x \cdot \mathcal{P}_{\bm{1}}^\perp \\
  &= \frac{Sn}{2} \frac{3}{N-1} \mathcal{P}_{\bm{1}}^\perp \nonumber
\end{align}
Although $Q_{S \leftarrow s}$ near the boundary (i.e. $S-s \ll n$) will not obey the approximate 1-design property (used to obtain the first equality),
  the sum of conjugated projection operators near the boundary in \eqnref{eq:W} only contributes at order $O(n^2)$,
  which is negligible compared to the other terms in \eqnref{eq:W2}.
We used $\Exp_U U_{ij} \otimes U_{kl}^* = \frac{1}{N} \delta_{ik} \delta_{jl}$ (where $\delta_{ij}$ denotes a Kronecker delta) to obtain the second line above.
The final line follows from the definitions of $\mathcal{P}_{\bm{1}}^\perp$ and $\mathcal{P}_x$ [\eqnref{eq:Px}].
This is the second term in \eqnref{eq:W2}.
To quantify statistical fluctuations about the above mean,
  we numerically find that the standard deviation of the eigenvalues of $\mathcal{W}$ within the subspace orthogonal to $\bm{1}$
  is approximately $4\sqrt{Sn/2N}$ [the third term in \eqnref{eq:W2}],
  which is much smaller than the $O(Sn/N)$ mean.

Note that $\bm{1} \cdot H = 0$ since the columns of $H$ sum to zero [\eqnref{eq:sumH0}].
Therefore inserting \eqnref{eq:W2} into \eqref{eq:dP2} yields
\begin{equation}
  \overline{\bm{P}^{(\tau)} - \bm{P}^{(\tau-1)}} \approx -\Delta_\text{t} iH \cdot \overline{\bm{P}^{(\tau-1)}} \label{eq:dP3}
\end{equation}
  up to $O(m_0^2,\delta_\text{t}^2,\Delta_\text{m}^2,\sqrt{N/S n})$ corrections, where
\begin{equation}
  \Delta_\text{t} = \frac{\Delta_\text{m}}{4} \, \frac{S n}{2 N} \frac{3}{1-N^{-1}} \, \delta_\text{t} \label{eq:Delta_t}
\end{equation}
is the effective EmQM time step.
Equation\,\eqref{eq:dP3} is precisely the discrete-time analog of the emergent Schr\"{o}dinger's equation \eqref{eq:SchroP} that we wanted.

The emergent wavefunction $\Psi(t)$ can be extracted from $\bm{P}^{(\tau)}$ in \eqnref{eq:P} as follows:
\begin{equation}
  \Psi(t = \Delta_\text{t} \tau) \propto \bm{P}^{(\tau)} - \frac{\bm{1}}{N} \label{eq:PsiEmQM}
\end{equation}
$\Psi$ obeys Schr\"{o}dinger's equation exactly in a
  $m_0,\delta_\text{t},S^{-1},\Delta_\text{m} \to 0$ limit,
  which we derive below.

\section{Deviations from Quantum Mechanics}
\label{sec:deviations}

We now study how much the EmQM model deviates from Schr\"{o}dinger's equation.
In particular, we will estimate how
\begin{equation}
  \varepsilon(t) = ||\Psi(t) - \Psi_\text{QM}(t)|| \label{eq:epst}
\end{equation}
  grows with time,
  where $\Psi$ is the EmQM wavefunction defined by \eqnref{eq:PsiEmQM},
  $\Psi_\text{QM}$ is calculated using Schr\"{o}dinger's equation \eqref{eq:Schro},
  and $||\cdots||$ denotes a Euclidean 2-norm.
At time $t=0$, we take $\Psi_\text{QM}(0) = \Psi(0)$.
The deviation $\varepsilon(t)$ from quantum mechanics increases due to four separate contributions resulting from finite
  $m_0$, $\delta_\text{t}$, $S^{-1}$, and $\Delta_\text{m}$.
We find that small values of these parameters result in
  linearity, unitarity, locality, and small statistical fluctuations, respectively.

Calculating the contributions to $\varepsilon(t)$ is rather technical.
For brevity, we will only sketch the derivation,
  which we verify numerically in \secref{sec:simulation}.
While estimating $\varepsilon(t)$, we also ignore constant factors (e.g. factors of 2),
  which we emphasize by using ``$\sim$'' symbols (instead of ``$\approx$'').

\subsection{Preliminaries}

It is essential to differentiate systematic and statistical errors,
  which respectively add coherently and incoherently.
That is, if one adds a series of $n$ errors following a normal distribution with
  mean $\mu$ (systematic error) and standard deviation $\sigma$ (statistical error),
  then the total error is also normally distributed with mean $n \mu$ and standard deviation $\sqrt{n} \sigma$.
We say that the means add coherently ($\propto n$) while the standard deviation adds incoherently ($\propto \sqrt{n}$).

It is instructive to first roughly estimate the 2-norm of $\overline{\bm{P}^{(\tau)} - \bm{P}^{(\tau-1)}} \approx - \Delta_\text{t} iH \cdot \bm{P}^{(\tau-1)}$ in \eqnref{eq:dP3}:
\begin{equation}
  \Delta_\text{P} \approx ||\Delta_\text{t} \, iH \cdot \bm{P}||
                  \sim \Delta_\text{t} \sqrt{n} \, \epsilon_\Psi \label{eq:DeltaP}
\end{equation}
This follows from $H \cdot \bm{P} = \epsilon_\Psi H \cdot \Psi$
  via the constraint \eqref{eq:sumH0} that $\sum_j H_{ij} = 0$.  
And $||H \cdot \Psi|| \sim \sqrt{n}$ due to the $n$ terms in $H = \sum_x H_x$ which add up incoherently for generically-random wavefunctions $\Psi$
  (which we consider in our numerical validation\footnote{%
  $||H \cdot \Psi|| \sim n$ for low-energy states.
  Such states thus require an extra factor of $\sqrt{n}$ in \eqnref{eq:DeltaP},
     but this is relatively negligible compared e.g. to factors of $N$.})
  due to approximate orthogonality.
This results in the $\sqrt{n}$ factor above.
For simplicity, we assume that each $H_x$ (viewed as a $4 \times 4$ matrix) has norm roughly equal to 1.

We will also require an estimate for $\epsilon_\Psi$,  which is defined by $\bm{P} = \frac{\bm{1}}{N} + \epsilon_\Psi \Psi$
  in \eqnref{eq:P} with $||\Psi|| = 1$:
\begin{equation}
  \epsilon_\Psi \sim \sqrt{\frac{S n}{N}} \, m_0 \label{eq:epsPsi}.
\end{equation}
This expression results from \eqnref{eq:p},
  which expresses $\bm{P} - \frac{\bm{1}}{N}$ as a sum of roughly $S n/2$ terms of the form $Q_{S \leftarrow s} \cdot m_{s,x} \cdot \frac{\bm{1}}{N}$.
These $S n/2$ terms add up incoherently, and each term has a norm of roughly $m_0 / \sqrt{N}$ (since $\bm{1}$ has norm $\sqrt{N}$),
  which results in the above expression.

\subsection{Small \texorpdfstring{$m_0$}{m0} Controls Linearity}

As noted in \secref{sec:linear},
  we only expect linearity to result if $M_{s,x}$ changes very little over time,
  which is controlled by the smallness of $m_0$.
This limit also allowed us to expand \eqnref{eq:p} to first order in $m_0$.
Keeping all higher-order terms yields a more accurate version of \eqnref{eq:dP}:
\begin{align}
  &\overline{\bm{P}^{(\tau)} - \bm{P}^{(\tau-1)}} = \\
    &\quad \sum_{s=1}^S \sum_x^{\substack{\text{even}\\s-x}}
      M_{S \leftarrow s} \cdot \left(\overline{m_{s,x}^{(\tau_s)} - m_{s,x}^{(\tau_s-1)}}\right) \cdot \frac{\bm{1}}{N} + O(\Delta_\text{m}^2) \nonumber
\end{align}
  where we only drop terms that are second order in
  $\Delta_\text{m} \sim \big|\big|m_{s,x}^{(\tau_s)} - m_{s,x}^{(\tau_s-1)}\big|\big|_\text{op} \ll m_0$.
We use $||M||_\text{op}$ to denote the operator norm of a matrix $M$,
  which is equivalent to the largest singular value of $M$.
Analogous to $Q_{S \leftarrow s}$,
  we define $M_{S \leftarrow s}$ 
  as a product of stochastic matrices $M_{s',x}$ with $S \geq s' > s$:
\begin{equation}
  M_{S \leftarrow s} = M_S \cdot M_{S-1} \cdots M_{s+1} \label{eq:MSs}
\end{equation}
This leads to a nonlinear Schr\"{o}dinger's equation
\begin{equation}
  \partial_t \Psi \approx -i \widetilde{\mathcal{W}} \cdot H \cdot \Psi \label{eq:nonlinearSchro}
\end{equation}
  where $\widetilde{\mathcal{W}}$ is similar to $\mathcal{W}$ in \eqnref{eq:W}
  but with $Q_{S \leftarrow s}$ replaced with $M_{S \leftarrow s}$:
\begin{equation}
  \widetilde{\mathcal{W}}
    = \sum_{s=1}^S \sum_x^{\substack{\text{even}\\s-x}}
        M_{S \leftarrow s} \cdot \mathcal{P}_x \cdot Q_{S \leftarrow s}^T \label{eq:Wtilde}
\end{equation}
\eqnref{eq:nonlinearSchro} is highly nonlinear because the right-hand-side depends on the product of many dynamical variables $m_{s,x}$ (via $\widetilde{\mathcal{W}}$) in addition to $\Psi$.
Furthermore, unlike Schr\"{o}dinger's equation,
  time evolving $\Psi$ using \eqnref{eq:nonlinearSchro} also requires keeping track of the time evolution of $m_{s,x}$.

To estimate how much nonlinear corrections to the emergent Schr\"{o}dinger equation contribute to the deviation $\varepsilon(t)$ in \eqnref{eq:epst},
  consider the $O(Sn)$ many $O(m_0)$ terms in $\widetilde{\mathcal{W}}$ that were neglected in $\mathcal{W}$.
These terms add up coherently\footnote{%
    There is also an incoherent contribution.
    However, the coherent contribution dominates for large $S$ since it adds up more rapidly.}
  by subtracting weight from the identity matrix component of $\widetilde{\mathcal{W}}$,
  such that
\begin{equation}
  \big|\big|\widetilde{\mathcal{W}} - \mathcal{W}\big|\big|_\text{op} \sim \min(S n m_0, 1) \label{eq:dW}
\end{equation}
The $\min(\cdots, 1)$ results because once $S n m_0 \gtrsim 1$,
  most of the weight has been removed from the identity matrix component.
This correction to $\mathcal{W}$ adds coherently to $\varepsilon(t)$ over many time steps.
Indeed, we find that $\partial_t \varepsilon(t) \propto \big|\big|\widetilde{\mathcal{W}} - \mathcal{W}\big|\big|_\text{op}$: 
\begin{align}
  \partial_t \varepsilon(t) &= \partial_t \, ||\Psi - \Psi_\text{QM}|| \nonumber\\
    &= \RE \frac{(\Psi - \Psi_\text{QM})^*}{||\Psi - \Psi_\text{QM}||} \cdot \partial_t (\Psi - \Psi_\text{QM}) \nonumber\\
    &\lesssim \left|\left| \IM \frac{(\Psi - \Psi_\text{QM})^*}{||\Psi - \Psi_\text{QM}||} \cdot (\widetilde{\mathcal{W}} \cdot H \cdot \Psi - \mathcal{W} \cdot H \cdot \Psi_\text{QM}) \right|\right| \label{eq:bound m0}\\
    &\sim \left|\left| \IM \frac{(\Psi - \Psi_\text{QM})^*}{||\Psi - \Psi_\text{QM}||} \cdot (\widetilde{\mathcal{W}} - \mathcal{W}) \cdot H \cdot \Psi \right|\right| \label{eq:bound m02}\\
    &\leq \big|\big|\widetilde{\mathcal{W}} - \mathcal{W}\big|\big|_\text{op} \; ||H \cdot \Psi|| \nonumber\\
    &\sim \min(Sn m_0,1) \sqrt{n} \label{eq:bound m03}
\end{align}
In \eqnref{eq:bound m0},
  we wish to bound how quickly the deviation $\varepsilon(t)$ increases just due to finite $m_0$.
Thus, we inserted the nonlinear \eqnref{eq:nonlinearSchro} in for $\partial_t \Psi$ and we used
  $\partial_t \Psi_\text{QM} \approx -i \mathcal{W} \cdot H \cdot \Psi_\text{QM}$
  since here we want to ignore the $O(\sqrt{Sn/N})$ corrections to $\mathcal{W}$ in \eqnref{eq:W2}.
Equation\,\eqref{eq:bound m02} also follows from ignoring the $O(\sqrt{Sn/N})$ corrections to $\mathcal{W}$ since
  $\IM (\Psi - \Psi_\text{QM})^* \cdot H \cdot (\Psi - \Psi_\text{QM}) = 0$.
\eqnref{eq:bound m03} follows from \eqnref{eq:dW}.
The $\sqrt{n}$ factor results from the $n$ terms in the Hamiltonian,
  which add incoherently as in \eqnref{eq:DeltaP}.

We therefore find that the small $m_0$ approximation contributes $\varepsilon_\text{m}(t)$ to $\varepsilon(t)$ where
\begin{equation}
  \varepsilon_\text{m}(t) \sim \min(Sn m_0,1) \sqrt{n} \, t \label{eq:eps m}
\end{equation}

\subsection{Small \texorpdfstring{$\delta_\text{t}$}{\textdelta t} Controls Unitarity}
\label{sec:unitarity}

To obtain \eqnref{eq:db}, which is inserted into \eqnref{eq:dmN 2}, we only kept terms with a single factor of $\delta_\text{t} H_x$, which is justified for small $\delta_\text{t}$.
Terms that are higher-order in $\delta_\text{t}$ lead to a non-unitary evolution in $\Psi$.
A more accurate version of \eqnref{eq:dP3} would replace
\begin{equation}
  -i \delta_\text{t} H \to \Otimes_x^\text{odd} B_x^{(+)} - \Otimes_x^\text{odd} B_x^{(-)} + \Otimes_x^\text{even} B_x^{(+)} - \Otimes_x^\text{even} B_x^{(-)} \label{eq:H to B}
\end{equation}
  which leads to an effective Hamiltonian that has $O(\delta_\text{t})$ non-Hermitian terms.

To estimate the contribution of finite $\delta_\text{t}$ to $\varepsilon(t)$,
  note that the right-hand-side of \eqnref{eq:H to B} contains roughly $n^2$ many $O(\delta_\text{t}^2)$ terms.
When \eqnref{eq:H to B} is inserted into Schr\"{o}dinger's equation,
  these $n^2$ terms add up incoherently,
  leading to an $O(n \delta_\text{t})$ correction.
Similar to \eqnref{eq:bound m03}, after many time steps,
  this correction adds coherently and contributes 
  $\varepsilon_\text{t}(t)$ to $\varepsilon(t)$, where
\begin{equation}
  \varepsilon_\text{t}(t) \sim n \delta_\text{t} \, t. \label{eq:eps t}
\end{equation}

\subsection{Large \texorpdfstring{$S$}{S} Controls Emergent Locality}
\label{sec:nonlocal}

\eqnref{eq:dP2} leads to a modified Schr\"{o}dinger's equation:
\begin{equation}
  \partial_t \Psi \approx -i \mathcal{W} \cdot H \cdot \Psi . \label{eq:SchroW}
\end{equation}
$\mathcal{W}$ was defined in \eqnref{eq:W} and consists of a sum of $S n/2$ projection matrices $Q_{S \leftarrow s} \cdot \mathcal{P}_x \cdot Q_{S \leftarrow s}^T$,
  which each project onto a 4-dimensional subspace.
These are the 4-dimensional subspaces for which the $Sn/2$ stochastic matrices $M_{s,x}$ can affect $\bm{P}$.
Therefore, $\mathcal{W}$ is a real symmetric matrix that has at most $2Sn$ non-zero eigenvalues.
If $2Sn \ll N$, then $\mathcal{W}$ reduces the dynamics of $\Psi$ to a random $(2Sn)$-dimensional subspace.\footnote{%
  The subspace is spanned by states that are
  each a symmetric superposition of a random quarter of the states in the classical basis,
    e.g. states like
    $\tfrac{1}{2} \ket{0011} + \tfrac{1}{2} \ket{0110} + \tfrac{1}{2} \ket{1010} + \tfrac{1}{2} \ket{1111}$.}
If $Sn \gg N$,
  then $\mathcal{W}$ has full rank;
  however, recall from \eqnref{eq:W2} that
  the eigenvalues of $\mathcal{W}$ are randomly distributed (due to the random $Q_{s,x}$)
  with a standard deviation that is $\sqrt{N/Sn}$ times smaller than the mean.
The eigenvectors of $\mathcal{W}$ do not have any special property that preserves locality.
Therefore, the dynamics of $\Psi$ are only local to the extent that $S n \gg N$.
That is, far-away terms in the effective Hamiltonian
  $\mathcal{W} \cdot H = \sum_x \mathcal{W} \cdot H_x$
  only commute in the $S n \gg N$ limit (which we have checked numerically);
  when $S n < N$, the noise in the randomness of $\mathcal{W}$ leads to nonlocal dynamics.

To intuitively understand the last point,
  once could consider approximating the randomness in $\mathcal{W} - \mathcal{W}_0$ as a symmetric matrix of small Gaussian random entries.
$\mathcal{W}_0$, defined in \eqnref{eq:W2}, is the non-random part of $\mathcal{W}$.
Since $\mathcal{W} - \mathcal{W}_0$ is highly-nonlocal,
  it makes $\mathcal{W} \cdot H$ also become nonlocal.
Technically, this nonlocality only occurs for circuit depths $S$ between $n \ll S \ll N/n$.
If the circuit depth is very small $S \ll n$,
  then the dynamics of the emergent wavefunction [\eqnref{eq:PsiEmQM}] will actually be local.
This results because the ``light-cone'' of back-propagating bits can only spatially extend to $\Delta x \sim S$ when $S \ll n$.
However, although the dynamics are local when $S \ll n$,
  $\mathcal{W}$ will have very few non-zero eigenvalues,
  which causes the EmQM model to be a very bad approximation to quantum mechanics.

The $S n \gg N$ requirement for locality is problematic because if there are many $n\gg1$ qubits,
  then the length $S$ of the extra dimension would have to be tremendously large
  ($S n \gg 2^n$) in order for $\Psi$ to have local dynamics.
Mitigating this nonlocality is an important future direction
  that we elaborate on in \secref{sec:modifications}.

Interestingly, $\mathcal{W} \cdot H$ at least has real eigenvalues.
This occurs because $\mathcal{W} \cdot H$ is similar\footnote{%
    Matrices $A$ and $B$ are similar if $B = P A P^{-1}$ for some invertible matrix $P$.}
  to
\begin{equation}
  \widetilde{H} = \mathcal{W}^{1/2} \cdot H \cdot \mathcal{W}^{1/2}
\end{equation}
  (when $\mathcal{W}$ is invertible),
  which implies that $\mathcal{W} \cdot H$ and $\widetilde{H}$ have the same eigenvalues.
$\mathcal{W}^{1/2}$ is Hermitian,
  which implies that $\widetilde{H}$ is also Hermitian.
Therefore $\widetilde{H}$ and $\mathcal{W} \cdot H$ both have real eigenvalues.
Furthermore, by absorbing $\mathcal{W}$ into the Hamiltonian as above,
  we obtain an effective Schr\"{o}dinger's equation:
\begin{equation}
  \partial_t \widetilde{\Psi} = -i \widetilde{H} \cdot \widetilde{\Psi}
\end{equation}
  where the wavefunction is
\begin{equation}
  \widetilde{\Psi} = \mathcal{W}^{-1/2} \cdot \Psi.
\end{equation}
However, the issue of nonlocality remains in the sense that $\widetilde{H}$ contains geometrically nonlocal and high-weight terms.

To estimate the contribution of finite $S$ to $\varepsilon(t)$,
  recall from \eqsref{eq:W2} that the eigenvalues of $\mathcal{W}$ are random with standard deviation $\sqrt{N/Sn}$.
This randomness induces an $O(\sqrt{N/Sn})$ correction to $\mathcal{W}$ in \eqnref{eq:SchroW}.
Similar to \eqnref{eq:bound m03}
  [but with $\widetilde{\mathcal{W}}$ and $\mathcal{W}$ respectively replaced by $\mathcal{W}$ and $\mathcal{W}_0$ from \eqnref{eq:W2}],
  this correction adds coherently over many time steps and contributes
  $\varepsilon_\text{S}(t)$ to $\varepsilon(t)$, where
\begin{equation}
  \varepsilon_\text{S}(t) \sim \sqrt{\frac{N}{S}} \, t. \label{eq:eps S}
\end{equation}
Similar to \eqsref{eq:DeltaP} and \eqref{eq:bound m03},
  the extra factor of $\sqrt{n}$ results from the $n$ terms in the Hamiltonian.

\subsection{Small \texorpdfstring{$\Delta_\text{m}$}{\textDelta m} Controls Statistical Fluctuations}

So far, we have only focused on the mean of $\bm{P}^{(\tau)}$.
But $\bm{P}$ is defined by the stochastic matrices $M_{s,x}$,
  which have stochastic dynamics that induce statistical fluctuations on the time evolution of $M_{s,x}$ and $\bm{P}$.
However, these statistical fluctuations will be small if the perturbations $m_{s,x}$
  change very slowly with time and aren't too small,
  which will give the output bits $a_S$ enough time to thoroughly sample the wavefunction $\Psi \propto \bm{P} - \bm{1}/N$.

The contribution of statistical fluctuations to $\varepsilon(t)$ can be estimated by
  considering how much $\bm{P}$ will be affected by statistical fluctuations after $\tau$ time steps.
$\bm{P}$ [defined in \eqnref{eq:Ps}] will only change after time steps for which an $M_{s,x}$ matrix changes,
  which only occurs if the $\pm$ back-propagating bits are different [see \eqnref{eq:dm}],
  i.e. when $b_{S,+\gamma} \neq b_{S,-\gamma}$ [defined below \eqnref{eq:dm2}].
For small $n \delta_t \ll 1$,
  this occurs for roughly a $n \delta_t$ fraction of time steps
  since $B_x^{(\pm)} = \mathbbl{1}_4 + O(\delta_\text{t})$ [\eqnref{eq:Bx0}].

We can then think of the statistical fluctuations as a random-walk in an $N$-dimensional space.
Recall that after $n_\text{steps}$ steps with typical step length $\ell_\text{step}$,
  a random walker will have moved a Euclidean distance of roughly $\ell_\text{step} \sqrt{n_\text{steps}}$.
After $\tau$ time steps, our random walker will move $n_\text{steps} \sim n \delta_\text{t} \tau$ times due to the arguments above.
Below, we argue that the length of each step will be roughly
\begin{equation}
  \ell_\text{step} \sim \frac{\Delta_\text{m}}{N} S n. \label{eq:l step}
\end{equation}
This implies that the statistical fluctuations to $\bm{P}$ will grow as
\begin{equation}
  \frac{\Delta_\text{m}}{N} S n \sqrt{n \delta_\text{t} \tau}. \label{eq:l walk}
\end{equation}

Consider a time step for which $b_{S,+\gamma} \neq b_{S,-\gamma}$.
Most of the $Sn/2$ perturbations $m_{s,x}$ will be modified by an amount proportional to $\Delta_\text{m}$ due to the resulting back-propagating bits.
Each such modification shifts $\bm{P}$ by roughly a distance $\Delta_\text{m}/\sqrt{N}$,
  since $N^{-1/2}$ is the norm of the uniform probability vector $\bm{1}/N$
  that multiplies $m_{s,x}$ in \eqnref{eq:p}.
However, one component of the shift to $\bm{P}$ adds up coherently (over the $Sn/2$ many perturbations $m_{s,x}$),
  while the other $N-1$ basis components add incoherently.
[The coherent component is spanned by $\bm{\hat{b}}_{S,+\gamma} - \bm{\hat{b}}_{S,-\gamma}$,
  which appears in \eqnref{eq:dmN 1}.]
The incoherent components are negligible when $Sn \gg N$.
Projecting onto the coherent contribution reduces $\Delta_\text{m}/\sqrt{N}$ by a factor\footnote{%
    To understand the $N^{-1/2}$ factor, consider the signed sum
      $\bm{S} = \sum_{k=1}^K \text{sign}\!\left(v^{(k)}_1\right) \bm{v}^{(k)}$
      of random unit-normalized length-$N$ vectors $\bm{v}^{(k)}$.
    Only the first component $S_1$ adds up coherently,
      while the other $N-1$ components add incoherently, resulting in
      $||\bm{S}|| \sim K/\sqrt{N} + \sqrt{K}$ for large $K$ and $N$.
    When $K\gg N$, the coherent contribution (first term) dominates.}
  of $N^{-1/2}$.
This explains the $\Delta_\text{m}/N$ factor in \eqnref{eq:l step}.
The $Sn$ factor occurs because there are $Sn/2$ many perturbations $m_{s,x}$.

These statistical fluctuations bound $\varepsilon(t) \geq \varepsilon_\text{stat}(t)$ where
\begin{equation}
  \varepsilon_\text{stat}(t) \sim \epsilon_\Psi^{-1} \frac{\Delta_\text{m}}{N} S n \sqrt{n \delta_\text{t} \tau} . \label{eq:stat0}
\end{equation}
The factor of $\epsilon_\Psi^{-1}$ [\eqnref{eq:epsPsi}] over \eqnref{eq:l walk} results from solving for $\Psi$ in \eqnref{eq:P}:
  $\bm{P} = \frac{\bm{1}}{N} + \epsilon_\Psi \Psi$.
The above simplifies to
\begin{equation}
  \varepsilon_\text{stat}(t) \sim m_0^{-1} \sqrt{\Delta_\text{m} n \, t} . \label{eq:stat}
\end{equation}
  after replacing $\tau \to t/\Delta_\text{t}$ using \eqnref{eq:Delta_t}.

\subsection{Negligible Delay}

There is an $O(S)$ discrete time delay between when a string of output bits $a_S$ is sampled
  to when the perturbations $m_{s,x}$ are updated.
However, if the perturbations $m_{s,x}$ change sufficiently slowly with time due to small $\Delta_\text{m}$,
  then this delay has a negligible effect on the wavefunction
  [which we will find to be the case in \eqnref{eq:delaySmall}].

To estimate the contribution to $\varepsilon(t)$,
  recall that in \eqnref{eq:dP2}
  we assumed that $\bm{P}^{(\tau)}$ varies slowly over $2S$ time steps such that
  $\bm{P}^{(\tau_s'-1)} = \bm{P}^{(\tau-1)} + O(S \Delta_\text{P})$ where $\tau_s' = \tau - 1 - 2(S-s)$, and $\Delta_\text{P}$ is defined in \eqnref{eq:DeltaP}.
This approximation induces an $O(\Delta_\text{t} \sqrt{n} S \Delta_\text{P})$ correction in \eqnref{eq:dP3},
  where the $\Delta_\text{t} \sqrt{n}$ follows for the same reason as in \eqnref{eq:DeltaP}.
After $\tau$ time steps, this correction adds coherently and contributes $\varepsilon_\text{delay}(t)$ to $\varepsilon(t)$, where
  $\varepsilon_\text{delay}(t) \sim \Delta_\text{t} \sqrt{n} S \Delta_\text{P} \, \tau / \epsilon_\Psi$,
  which simplifies to
\begin{equation}
\begin{aligned}
  \varepsilon_\text{delay}(t) &\sim \Delta_t n S \, t \\
    &\sim \Delta_\text{m} \frac{n^2 S^2}{N} \delta_\text{t} \, t.
\end{aligned} \label{eq:epsLag}
\end{equation}

\subsection{Convenient Limit}

The above contributions to $\varepsilon(t) = ||\Psi(t) - \Psi_\text{QM}(t)||$ add up incoherently, such that the total deviation from Schr\"{o}dinger's equation is roughly
\begin{equation}
   \varepsilon(t)
     \sim \sqrt{\varepsilon_\text{m}(t)^2 + \varepsilon_\text{t}(t)^2 + \varepsilon_\text{S}(t)^2 + \varepsilon_\text{stat}(t)^2} \label{eq:epsMax}
\end{equation}
until saturation near orthogonality at $\varepsilon(t) \approx \sqrt{2}$.
$\varepsilon_\text{delay}(t)$ also contributes,
  but we neglect it here since it contributes negligibly in the limit that we consider.

It is convenient to consider a limit of $S$, $\delta_\text{t}$, $m_0$, and $\Delta_\text{m}$
  as a function of a single small parameter $\epsilon_0$
  such that $\varepsilon(t) \to 0$ in the $\epsilon_0 \to 0$ limit.
We shall consider the parameterization such that all errors are roughly equal at time $t=\epsilon_0^{-1}$:
\begin{equation}
\begin{aligned}
  \varepsilon_\text{m}(t) &\sim \varepsilon_\text{t}(t) \sim \varepsilon_\text{S}(t) \sim \epsilon_0 \, t \\
  \varepsilon_\text{stat}(t) &\sim \sqrt{\epsilon_0 \, t}
\end{aligned} \label{eq:epsEq}
\end{equation}
This parameterization results in a deviation
\begin{align}
   \varepsilon(t) &\sim \sqrt{\epsilon_0 t + 3(\epsilon_0 t)^2} \label{eq:err} \\
      &\sim \sqrt{\epsilon_0 t} \text{ when  } t \lesssim t_0^{-1} \nonumber
\end{align}
  which is dominated by statistical fluctuations $\varepsilon_\text{stat}(t) \sim \sqrt{\epsilon_0 t}$ until $t\geq \epsilon_0^{-1}$.
Solving \eqnref{eq:epsEq} for
  $S$, $\delta_\text{t}$, $m_0$, and $\Delta_\text{m}$
  results in the following parameterization:
\begin{equation}
\begin{aligned}
  S &\approx N \, \epsilon_0^{-2} \\
  \delta_\text{t} &= \frac{\epsilon_0}{n} \\
  m_0 &= \frac{\epsilon_0}{S n^{3/2}} \approx \frac{\epsilon_0^3}{n^{3/2} N} \\
  \Delta_\text{m} &= \frac{m_0^2 \epsilon_0}{n} \approx \frac{\epsilon_0^7}{n^4 N^2}
\end{aligned} \label{eq:epsilon}
\end{equation}

Plugging the above parameters into $\varepsilon_\text{delay}(t)$ in \eqnref{eq:epsLag}
  shows that deviations due to $\varepsilon_\text{delay}(t)$ are negligibly small:
\begin{equation}
  \varepsilon_\text{delay}(t) \sim \frac{\epsilon_0^4}{n^3 N} \, t \ll \varepsilon(t) \label{eq:delaySmall}
\end{equation}

\section{Simulation}
\label{sec:simulation}

\begin{figure*}[t!]
  \centering
  \subfloat[$n=4$ \label{fig:fidelity4}]{\includegraphics[width=.9\columnwidth]{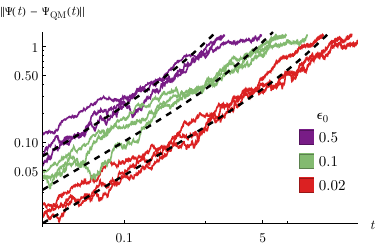}} \hspace{.15\columnwidth}
  \subfloat[$n=6$]{\includegraphics[width=.9\columnwidth]{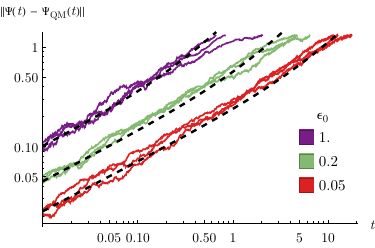}} \\
  \subfloat[$n=4$]{\includegraphics[width=.9\columnwidth]{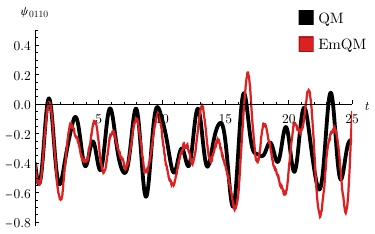}} \hspace{.15\columnwidth}
  \subfloat[$n=6$]{\includegraphics[width=.9\columnwidth]{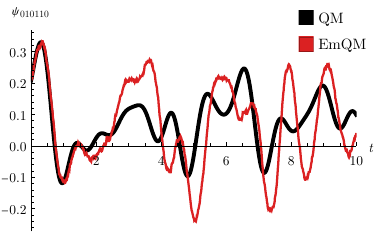}} \\
  \caption{%
    {\bf (a-b)} To validate our error estimates, we plot the deviation $\varepsilon(t) = ||\Psi(t) - \Psi_\text{QM}(t)||$ vs time $t$ 
        between the quantum mechanics (QM) wavefunction $\Psi_\text{QM}$
        and emergent QM (EmQM) wavefunction $\Psi$ [\eqnref{eq:PsiEmQM}] for (a) $n=4$ and (b) $n=6$ qubits
        using three different random initializations (colored lines)
        for each choice of the control parameter $\epsilon_0$ [see \eqnref{eq:epsilon}].
    As $\epsilon_0$ decreases, the EmQM model becomes increasingly more accurate, roughly agreeing with QM out to time $t \sim \epsilon_0^{-1}$.
    Dashed black lines show the estimated deviation $\varepsilon(t) \sim \sqrt{\epsilon_0 t + 3(\epsilon_0 t)^2}$ [\eqnref{eq:err}] for the three different values of $\epsilon_0$,
      which match the simulated data remarkably well.
    {\bf (c-d)} To demonstrate that the EmQM model is reproducing nontrivial dynamics, we also plot the $0110$ (and $010110$) component of the
      $n=4$ (and $n=6$) wavefunctions vs time
      for QM and the EmQM model with $\epsilon_0=0.02$ (and $\epsilon_0=0.05$), for which $S=800$ (and $S=1280$) [in accordance with \eqnref{eq:epsilon}].
  }\label{fig:fidelity}
\end{figure*}

\begin{figure}[t!]
  \centering
  \includegraphics[width=\columnwidth]{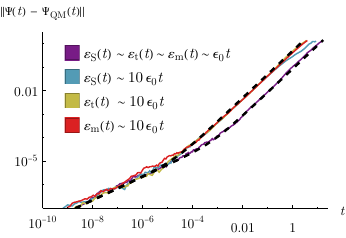}
  \caption{%
    The deviation $\varepsilon(t) = ||\Psi(t) - \Psi_\text{QM}(t)||$ from quantum mechanics (QM) vs time $t$.
    The deviation is dominated by statistical fluctuations $\varepsilon_\text{stat}(t)$ until the change in slope,
      after which the deviation is dominated by other contributions.
    Blue, yellow, and red lines show dominant contributions from $\varepsilon_\text{S}(t)$, $\varepsilon_\text{t}(t)$, and $\varepsilon_\text{S}(m)$, respectively defined in the legend
      (with $\varepsilon_\text{stat}(t) \sim 10^{-2} \sqrt{\epsilon_0 t}$ and $\varepsilon_\text{S}(t) \sim \varepsilon_\text{t}(t) \sim \varepsilon_\text{m}(t) \sim \epsilon_0 t$ whenever unspecified).
    Simulations are for $n=4$ qubits and $\epsilon_0 = 0.05$
      with parameters $S$, $\delta_\text{t}$, $m_0$, and $\Delta_\text{m}$
      chosen to target the previously-mentioned contributions
      [using \eqsref{eq:eps m}, \eqref{eq:eps t}, and \eqref{eq:eps S}].
    The dashed lines plot the expected deviations $\varepsilon(t)$ using \eqnref{eq:epsMax},
      which agree remarkably well with the simulation data (colored lines).
  }\label{fig:fidelityErrors}
\end{figure}

To numerically verify our theoretical results,
  we simulate the Hamiltonian $H = \sum_x H_x$ within the EmQM model with
\begin{equation}
  H_x = Y_x X_{x+1} - Y_x \label{eq:Hx}
\end{equation}
  which is a simple choice that satisfies the Hamiltonian constraints \eqref{eq:sumH0}.
To do this, we pick a $\epsilon_0$ to define the model parameters according to \eqnref{eq:epsilon}.
We then initialize the $m_{s,x}$ matrices using Gaussian random numbers with standard deviation $m_0$
  and then subtract a constant from each column such that all columns sum to zero.
This random initialization implicitly defines a random wavefunction [via \eqref{eq:PsiEmQM}].
We then time evolve the circuit for many steps.
$\Psi$ is normalized and extracted from $\bm{P}$ using \eqnref{eq:PsiEmQM}.

However, simulating the EmQM model is extremely expensive for small $\epsilon_0$ or many qubits since the CPU time
  required to simulate out to time $t$ scales as
\begin{equation}
  \text{CPU time} \sim S n \, t/\Delta_\text{t} \sim \frac{N}{\delta_\text{t} \Delta_\text{m}} t \approx \frac{n^5 N^3}{\epsilon_0^8} \, t \label{eq:CPU}
\end{equation}
The first relation follows since $t/\Delta_\text{t}$ time steps are required on an $S \times n$ lattice.
The second relation is obtained by inserting \eqnref{eq:Delta_t} for $\Delta_\text{t}$.
The final expression is valid for the $\epsilon_0$ parameterization \eqref{eq:epsilon}.
However, we can approximately simulate the EmQM model with high accuracy using the significantly-faster method defined and verified in \appref{app:approximate}.

We validate our theory for the EmQM model in \figref{fig:fidelity} by plotting the deviation $\varepsilon(t)$
  of the emergent wavefunction $\Psi$ from the quantum mechanics prediction.
We find that the deviation is in agreement with the estimated \eqnref{eq:err}.
This verifies that as the control parameter $\epsilon_0$ decreases,
  the EmQM model deviates less from quantum mechanics.
All data is generated using the approximate simulation method (\appref{app:approximate})
  except for the $n=4$ data with $\epsilon_0 = 0.5$ (purple in \figref{fig:fidelity4}),
  for which we could directly simulate the EmQM model.

The deviations from quantum mechanics shown in \figref{fig:fidelity} are dominated by the statistical deviation $\varepsilon_\text{stat}(t)$.
To verify the other contributions to the deviation $\varepsilon(t)$,
  we solve for parameters $S$, $\delta_\text{t}$, $m_0$, and $\Delta_\text{m}$
  such that the statistical deviation is much smaller and such that only one of the other contributions is expected to dominate.
This allows us to verify each contribution individually in \figref{fig:fidelityErrors}.

\section{Experimental Signatures}
\label{sec:experiment}

\subsection{Many Entangled High-Fidelity Qubits}
\label{sec:fidelityExperiment}

Evidence that an EmQM model really describes nature might be found by measuring deviations from Schr\"{o}dinger's equation,
  such as the $\varepsilon(t)$ [\eqnref{eq:epst}] studied in the previous section.
In \secref{sec:deviations},
  we calculated several contributions to $\varepsilon(t)$.
All contributions to $\varepsilon(t)$ can be made extremely small for very long times by taking $S^{-1}$, $\delta_\text{t}$, $m_0$, and $\Delta_\text{m}$ to be very small.
$\varepsilon_\text{S}(t) \sim t \sqrt{N/S}$ in \eqnref{eq:eps S} is the only deviation that increases polynomially with $N$.
This increase with $N$ is significant because $N=2^n$ is exponentially large in the number of qubits $n$.
Therefore, even if $S \sim 10^{100}$ or $10^{1000}$,
  only $n \sim 350$ or $3500$ qubits would be needed to obtain a large deviation
  $\varepsilon_\text{S}(t) \sim 1$ after a short time $t \lesssim 1$.

But what should be the value of $n$?
If $n$ is taken to be the number of qubits needed to describe just a mesoscopic region of space,
  e.g. Avogadro’s number $n \sim 10^{23}$,
  then a large deviation from quantum mechanics due to $\varepsilon_\text{S}(t)$ is already predicted for very short times unless $S$ is extremely large $S \gg 2^{10^{23}}$.
Indeed, in \secref{sec:nonlocal} we found that our model exhibits nonlocal EmQM dynamics unless $S n \gg 2^n$.
Therefore, it seems implausible that the particular EmQM model studied in this work
  accurately describes a possible EmQM for our universe.

However, we speculate that modifications to our model,
  such as those discussed in \secref{sec:modifications},
  could alleviate the $S n \gg 2^n$ requirement for local dynamics of the emergent wavefunction and thus
  yield a more useful toy model for EmQM.
For example,
  perhaps deviations from quantum mechanics might only
  be detectable if $S n \gg 2^{\widetilde{n}}$,
  where $\widetilde{n}$ is
  the number of highly entangled qubits that are measured with high fidelity.
This hypothesis has not yet been tested experimentally beyond very modest values of $\widetilde{n}$, but might be tested in the future for gradually increasing values of $\widetilde{n}$ by executing deep quantum circuits using quantum computers. Such experiments would significantly constrain the length $S$ of the extra dimension because of the requirement that $S$ be exponential in $\widetilde{n}$.
This idea motivated \refcite{SlagleTesting},
  which proposed to test the validity of quantum mechanics
  using a Loschmidt echo circuit on many qubits.

\subsection{Bell Inequality Tests}

Any attempt to describe quantum reality in terms of an underlying local classical model faces the potential obstacle that locally realistic classical models conform to Bell inequalities which are known to be experimentally violated. Yet our model of EmQM agrees with quantum mechanics to high accuracy and so can be expected to pass such tests. 

One way to understand why Bell inequality violation cannot easily exclude our model is to note that the speed of information propagation among the underlying classical bits, though finite, is much faster than 
the emergent speed of light on the boundary. This feature makes it exceedingly hard to close the ``locality loophole'' --- that is, to rule out communication between Alice's and Bob's labs during the test. 

The classical bits carry information at the speed
\begin{equation}
  v_\text{fast} \approx \frac{l_0}{\Delta_\text{t}} \approx \frac{l_0}{\Delta_m \delta_\text{t}} \frac{N}{S n}
\end{equation}
  where $l_0$ is the spatial distance between bits and
  where $\Delta_\text{t}$ [\eqnref{eq:Delta_t}] is how much time elapses in the EmQM for each discrete time step.
If, for example, we insert the $\epsilon_0$ parameterization from \eqnref{eq:epsilon} and neglect negligible factors of $n$, we obtain:
\begin{equation}
  v_\text{fast} \sim \frac{l_0}{t_0} N^2 \epsilon_0^{-6}
\end{equation}
  where $t_0^{-1} \sim ||H_x||_\text{op}$ is the norm of the local terms in the Hamiltonian
  (which we previously set to be roughly equal to 1).
However, according to local quantum mechanics, all particle velocities (e.g. the speed of light)
  should be upper bounded by $v_\text{QM} \sim l_0 / t_0$.
Therefore, since $v_\text{fast} \gg v_\text{QM} \geq c$,
  Bell tests cannot easily detect signatures of our model.

\section{Outlook}
\label{sec:outlook}

In future work, it will be interesting to investigate how generic emergent quantum mechanics (EmQM) is.
That is, if our model is changed slightly,
  will EmQM still be exhibited?
Or do additional ingredients need to be added to our model such that
  EmQM is a generic result?
Or from another point of view,
  can EmQM be thought of as a highly-exotic phase of classical matter?
In a sense,
  research showing that quantum mechanics is ``an island in theory space'' \cite{AaronsonTheoryspaceIsland}
  that can be derived axiomatically \cite{HardyAxioms,KapustinQM,QuantumFromPrinciples}
  suggests that EmQM might indeed be a stable fixed point under coarse-graining in a broad class of local classical models.
It would also be useful to determine if EmQM can result without relying on very small parameters (e.g. $\delta_\text{t}$, $m_0$, and $\Delta_\text{m}$).

\subsection{Mitigating Nonlocality}
\label{sec:modifications}

As emphasized in \secref{sec:fidelityExperiment},
  a crucial remaining future direction is to determine if
  modifications of our EmQM model could alleviate the
  $Sn \gg 2^n$ requirement for local EmQM dynamics.
We would prefer to have an EmQM model such that $S \gg 2^{\widetilde{n}}$ implies consistency with any experiment that only probes $\widetilde{n}$ highly entangled qubits with high fidelity, e.g. the logical qubits in a quantum computer.
This would be desirable because only $\widetilde{n} \sim \log_2 S$ highly entangled qubits would be needed to experimentally test such a model of EmQM,
  which would be experimentally relevant in the near-term if e.g. $S \sim 2^{1000}$.

Nonlocal dynamics when $Sn \ll 2^n$ in our model may result because the stochastic circuits we consider are not very efficient at encoding the wavefunction.
In particular, the emergent wavefunction is encoded using random permutation matrices.
This inefficient encoding could be contrasted with MERA tensor networks \cite{MERA} or deep neural networks,
  where each layer can perform a more useful entanglement renormalization \cite{TNRMERA} or coarse graining \cite{deepRG}.

One possible approach to achieve more efficient circuits could be to relax the requirement \eqref{eq:MQm} that
  the stochastic matrices $M_{s,x}$ are perturbatively close to permutation matrices.
But then the subleading (i.e. all but the largest) singular values of $M_{S \leftarrow s}$ [\eqnref{eq:MSs}] will generically be exponentially small in $S-s$.
If that occurs, then the overwhelming majority of singular values of $\widetilde{\mathcal{W}}$ [\eqnref{eq:Wtilde}] will also be extremely small,
  which will lead to emergent dynamics [\eqnref{eq:nonlinearSchro}] that do not approximate quantum mechanics well.
In this scenario, deep circuits of stochastic matrices are not useful because each layer destroys too much information.

To mitigate this problem,
  we could consider promoting the classical bits to real numbers.
Then the permutation matrices of two bits are promoted to invertible functions from $\mathbb{R}^2$ to $\mathbb{R}^2$,
  which can be viewed as a permutation of $\mathbb{R}^2$.
But unlike permutation matrices, such functions can map the uniform distribution to a different probability distribution.
Furthermore, this map can be perfectly inverted.
Therefore, unlike deep circuits of generic stochastic matrices,
  deep circuits of functions do not destroy information.
Composing a deep circuit of functions in this way can produce arbitrary probability distributions
  (and thus arbitrary wavefunctions for EmQM), an observation
  which has been utilized within the deep learning community \cite{NormalizingFlows,FFJORD}.

\subsection{Quantum Computation and Fundamental Physics}

If quantum mechanics does emerge from classical mechanics,
  then the computational power of quantum computers could be severely limited \cite{HooftBook,SlagleTesting}.
For example, BQP-hard problems may only be tractable in  actual devices for limited problem sizes.
On the other hand, it is possible that deviations from quantum mechanics (such as nonlinear corrections to the Schr\"{o}dinger equation) could enhance the power of quantum computers
  \cite{AbramsNonlinearNP,AaronsonPostBQP} for some problems of (possibly) limited size.

Even more speculatively, discovering that quantum mechanics emerges from an underlying local classical model  might open new directions for understanding dark matter, dark energy, early-universe cosmology, and the black hole information paradox \cite{QuantumGravityNecessary}.
Finally, that our EmQM model encodes the quantum wavefunction on the boundary of an extra spatial dimension suggests possible connections to holographic duality in quantum gravity \cite{AdSCFT}.

\begin{acknowledgments}
We thank Jacques Pienaar, Scott Aaronson, Xie Chen, Jason Alicea, Monica Kang, and Stefan Prohazka
  for valuable discussions.
K.S. was supported by the Walter Burke Institute for Theoretical Physics at Caltech; and
  the U.S. Department of Energy, Office of Science, National Quantum Information Science Research Centers, Quantum Science Center.
J.P. acknowledges funding provided by the Institute for Quantum Information and Matter, an NSF Physics Frontiers Center (PHY-1733907), the Simons Foundation It from Qubit Collaboration, the DOE QuantISED program (DE-SC0018407), and the Air Force Office of Scientific Research (FA9550-19-1-0360).
\end{acknowledgments}

\bibliography{EmQM}

\appendix

\section{Measurements}
\label{app:measurement}

In this appendix, we clarify how measurements could be interpreted in our model within the Everett interpretation of quantum theory.
We also speculate how possible improvements upon our EmQM model might lead to a resolution of the measurement problem.

\subsection{Everett Interpretation Review}

In the Everett interpretation,
    the observer is included in the wavefunction.
In principle, the measurement process can then be formalized as a Hamiltonian evolution via Schr\"odinger's equation.
After the measurement, Schr\"odinger's equation predicts that the observer becomes entangled with the measured system.
That is, the resulting wavefunction is in a superposition of states,
    where each state describes one of the measurement outcomes (from the observer's perspective).

For example, consider an observer who measures whether the state of a spin is $\ket{\uparrow}$ or $\ket{\downarrow}$.
Before the measurement, suppose that wavefunction of the spin is
\begin{equation}
  \ket{\psi_\text{before}} = a \ket{\uparrow} + b \ket{\downarrow}
\end{equation}
In the Everett interpretation, we imagine describing the measurement process using a wavefunction $\ket{\Psi}$ for the entire universe.
Before the measurement, we schematically write the universe's wavefunction as
\begin{equation}
  \ket{\Psi_\text{before}} = \ket{\mathsf{observer}} \otimes (a \ket{\uparrow} + b \ket{\downarrow})
\end{equation}
  where $\ket{\mathsf{observer}}$ is the wavefunction for the observer (and the rest of the universe other than the spin).
After the measurement (assuming it is performed perfectly),
  the wavefunction of the universe should be
\begin{equation}
\begin{aligned}
  \ket{\Psi_\text{after}} &= a \ket{\mathsf{observer\ sees }\uparrow} \otimes \ket{\uparrow} \\
             &+\, b \ket{\mathsf{observer\ sees }\downarrow} \otimes \ket{\downarrow}
\end{aligned} \label{eq:worlds}
\end{equation}
  where $\ket{\mathsf{observer\ sees }\uparrow} \otimes \ket{\uparrow}$
  is the wavefunction of the universe where the observer has observed the $\ket{\uparrow}$ state,
  and similar for $\ket{\mathsf{observer\ sees }\downarrow} \otimes \ket{\downarrow}$.

This exemplifies that in the Everett interpretation,
  the wavefunction consists of a superposition of all measurement outcomes.
The different states in the superposition are macroscopically different
  and are consequently extremely unlikely to significantly interfere with each other.
These different states effectively behave as different ``worlds.''
Indeed, this interpretation is also often referred to as the \emph{many-worlds interpretation}.

But then how does one predict the probability that an observer will measure a given outcome (from the observer's perspective)?
Born's rule implies that the probabilities for the two states, or worlds,
  in \eqnref{eq:worlds} are $|a|^2$ and $|b|^2$.
But in the Everett interpretation, there are no measurement axioms or applications of Born's rule;
  only Schr\"odinger's equation is used to evolve the wavefunction.
In Schr\"odinger's equation,
  $a$ and $b$ are just coefficients in a linear expansion,
  and it is not clear why or how these coefficients should be assigned to probabilities associated with the observer's experience.

Nevertheless, many authors have argued \cite{DeutschBorn,WalliceBorn,CarrollBorn,ZurekBorn,MasanesBorn,HossenfelderBorn,bookBorn}
  that Born's rule is the only reasonable or consistent choice,
  under various reasonable assumptions.
For example, \refcite{HossenfelderBorn} gives a brief argument that merely assumes unitary invariance, continuity, and system size invariance.
Yet, it remains controversial whether the Everett interpretation (without measurement axioms) provides a complete description of quantum theory \cite{KentAgainstManyWorlds}.

\subsection{EmQM and the Everett Interpretation}

In our EmQM model, the slow time dynamics of the boundary degrees of freedom can be accurately predicted using Schr\"odinger's equation.
That is, there is an emergent wavefunction [$\Psi$ in \eqnref{eq:P}]
  whose time dynamics can be well-approximated by Schr\"odinger's equation.
Similar to Everett's interpretation,
 neither measurement axioms nor Born's rule explicitly appear in our EmQM model.
Thus we are left with a similar challenge as in Everett's interpretation:
  does Born's rule somehow follow from the equations of motion alone?

But for a model of EmQM from classical mechanics,
  we would not expect the emergent wavefunction to be able to store very many different worlds in superposition.
Instead, there must eventually be some sort of ``collapse'' of the emergent wavefunction to only a subset of the superposed measurement outcomes.
Importantly, this collapse of states should be (at least approximately) consistent with Born's rule if the EmQM model closely approximates standard quantum theory.
Perhaps an EmQM model with this property could provide a satisfying solution to the measurement problem.

Unfortunately, our EmQM model does not appear to be successful enough to study this hypothetical wavefunction collapse.
As explained in \secref{sec:fidelityExperiment},
  in a certain limit with an extremely large extra dimension,
  our model simply reproduces Schr\"odinger's equation.
Therefore this limit is similar to the Everett interpretation,
  and it is not obvious how the measurement probabilities for an observer should be assigned,
  although some of the arguments in \refscite{DeutschBorn,WalliceBorn,CarrollBorn,ZurekBorn,MasanesBorn,HossenfelderBorn,bookBorn}
  might still be applicable. 
If the extra dimension isn't sufficiently large,
  then there are nonlocal violations of Schr\"odinger's equation, but we have no reason to expect these violations to imply an approximate version of Born's rule
  (although we have not checked thoroughly).
It seems that better models are needed to assess whether the Born rule, in addition to Schr\"{o}dinger's equation, could arise from a sensible classical model of EmQM.

\section{Real-valued Quantum Mechanics}
\label{app:realQM}

In this appendix, we show that any quantum Hamiltonian and wavefunction in Schr\"odinger's equation can be linearly mapped to real-valued analogs with zero row and column sums that satisfy Eqs.\,(\ref{eq:sumPsi}--\ref{eq:sumH0}) while preserving locality.
We do this by first mapping to real-valued quantum mechanics \cite{McKagueRealQM,AleksandrovaRealQM,StueckelbergRealQM,MyrheimRealQM}
  in \appref{app:real} and then focus on zero sums in \appref{app:zero}.

Both mappings generically require adding additional qubits.
Preserving locality requires multiplying the qubit count by a constant factor (when using the systematic mapping).
As a result, the possible real-valued wavefunctions that result from this mapping are highly constrained in the sense that these wavefunctions only span a subset of the Hilbert space.
Therefore, although there are well-known fundamental differences between complex and real-valued quantum mechanics \cite{AaronsonTheoryspaceIsland,RealQMFalsifiable,RealQMExp,RealQMExperiment,HardyAxioms,CavesRebits,CavesFinetti},
  this mapping shows that complex-valued quantum mechanics is equivalent to real-valued quantum mechanics
  constrained to a \emph{subspace} of the Hilbert space.
However in \appref{app:realDynamics},
  we emphasize that in real-valued quantum mechanics,
  it is not correct to assume that ``the state representing two independent preparations of the two systems is the tensor product of the two preparations'' \cite{RealQMFalsifiable}.

\subsection{Mapping from Complex to Real QM}
\label{app:real}

\subsubsection{Geometrically nonlocal Mapping}

If we do not require that the mapping is geometrically local,
  then mapping to real values can be achieved simply by splitting complex numbers into their real and imaginary parts \cite{McKagueRealQM,AleksandrovaRealQM,   StueckelbergRealQM}.
This can be achieved for operators via the replacement
\begin{equation}
  i \to -i \sigma^2 \label{eq:i}
\end{equation}
  where 
  \begin{equation}
  \sigma^2 =\begin{pmatrix}0&-i\\i&0 \end{pmatrix}
  \end{equation}
  is a Pauli operator acting on an additional qubit.

We note that 
\begin{equation}
  \ket{\pm i} = \tfrac{1}{\sqrt{2}} \big( \ket{\uparrow} \pm i\ket{\downarrow} \big).
\end{equation}
are eigenstates of $\sigma^2$ with eigenvalues $\pm 1$, and that 
\begin{equation}
    P(\pm) = \tfrac{1}{2}\left(\mathbbl{1}\pm\sigma^2\right)
\end{equation}
are orthogonal projectors onto these eigenstates. Furthermore, $\ket{+i}$ and $\ket{-i}$ are complex conjugates of one another, as are $P(+)$ and $P(-)$. We map an $n$-qubit wavefunction $\ket{\psi}$ to a real $(n{+}1)$-qubit wavefunction $\ket{\widetilde{\psi}}$ according to
\begin{align}
  \ket{\psi} \to \ket{\widetilde{\psi}} =\tfrac{1}{\sqrt{2}} \big( \ket{\psi} \otimes \ket{-i} + \ket{\psi}^* \otimes \ket{+i} \big),
  \end{align}
  and map an $n$-qubit operator $Q$ to a real $(n{+}1)$-qubit operator $\widetilde{Q}$ according to 
  \begin{align}
  Q \to \widetilde{Q} = Q \otimes P(-) + Q^*\otimes P(+)\label{eq:QMap}
\end{align}
where * denotes complex conjugation; thus
\begin{align}
    \widetilde{Q} \ket{\widetilde{\psi}} = \tfrac{1}{\sqrt{2}} \big( Q\ket{\psi} \otimes \ket{-i} + Q^*\ket{\psi}^* \otimes \ket{+i} \big).
\end{align}
Under this mapping, Schr\"{o}dinger's equation
  $\partial_t \ket{\psi(t)} = -iH \ket{\psi(t)}$ is equivalent to
  \begin{align}
  \partial_t \ket{\widetilde\psi(t)} =- (\widetilde{iH})\ket{\widetilde\psi(t)}.
  \end{align}
That is, it is mapped to Schr\"{o}dinger's equation for the real wavefunction $\ket{\widetilde{\psi}}$ with imaginary Hamiltonian $-i(\widetilde{iH})$,
  where $\widetilde{iH}$ denotes the result of mapping $Q=iH$ using \eqnref{eq:QMap}.

Another way to express the mapping is sometimes convenient. Note that if $Q$ is real, then 
 \begin{equation}
 \widetilde{Q} = Q\otimes \mathbbl{1},
 \end{equation}
 while if $Q$ is imaginary, then
 \begin{equation}
 \widetilde{Q} = (-iQ)\otimes (-i\sigma^2);
 \end{equation}
 more generally
 \begin{equation}
     \widetilde{Q} = \RE Q\otimes \mathbbl{1}+ \IM Q\otimes (-i\sigma^2).
 \end{equation}
Similarly, if $\ket{\psi}$ is real in some basis, then
 \begin{equation}
 \ket{\widetilde{\psi}} = \ket{\psi} \otimes \ket{\uparrow},
 \end{equation}
 while if $\ket{\psi}$ is imaginary, then
 \begin{equation}
 \ket{\widetilde{\psi}} = -i\ket{\psi} \otimes \ket{\downarrow};
 \end{equation}
 and more generally
 \begin{equation}
 \ket{\widetilde{\psi}} = \ket{\RE\psi} \otimes \ket{\uparrow} + \ket{\IM\psi} \otimes \ket{\downarrow}.
 \end{equation}

\subsubsection{Local Mapping}

Now suppose that $H = \sum_x H_x$ is a local Hamiltonian, where each $H_x$ only acts on qubits near the spatial point $x$. If $H_x$ and $H_y$ act on sets of qubits that are distantly separated from one another, then it's not possible to put the additional qubit close to both sets. Therefore, the above operator map does not preserve the geometric locality of the Hamiltonian. 

In order to promote the nonlocal mapping to a mapping that preserves locality for local operators $Q_x$,
  we can instead add a new
  qubit adjacent to each lattice site 
  and map
\begin{align}
  i \to -i \sigma_x^2 \label{eq:ix}
\end{align}
at each site.
Wavefunctions are then mapped according to
\begin{align}
\ket{\psi} &\to \ket{\widetilde{\psi}} =\tfrac{1}{\sqrt{2}} \big( \ket{\psi} \Otimes_x \ket{-i}_x + \ket{\psi}^* \Otimes_x \ket{+i}_x \big). \label{eq:localWavefunctionMap}
\end{align}
Now suppose that $Q_x$ is a $k$-local operator (e.g. a term in the Hamiltonian) acting on $k$ qubits near $x$. After the mapping, the site $x$ is accompanied by an adjacent auxiliary qubit, and $Q_x$ is mapped to
\begin{align}
    \widetilde{Q}_x = Q_x \otimes P(-)_x + Q_x^*\otimes P(+)_x,
\end{align}
a $(k+1)$-local operator acting on the $k$ qubits together with the auxiliary qubit adjacent to site $x$.
Another way to express the mapping is
\begin{align}
    \widetilde{Q}_x = \RE Q_x \otimes \mathbbl{1}_x + \IM Q_x \otimes (-i\sigma_x^2).
\end{align}
Thus the $k$-local Hamiltonian $H = \sum_x H_x$ can be mapped to the $(k+1)$-local Hamiltonian $\widetilde{H} = \sum_x\widetilde{H_x}$.

As we found for the nonlocal mapping, Schr\"{o}dinger's equation 
  $\partial_t \ket{\psi(t)} = -iH \ket{\psi(t)}$ is equivalent to
  to Schr\"{o}dinger's equation for the real wavefunction $\ket{\widetilde{\psi}}$ with imaginary Hamiltonian
\begin{equation}
  -i(\widetilde{iH}) =\sum_x\left( i \IM (H_x) \otimes \mathbbl{1}_x - \RE (H_x) \otimes \sigma_x^2\right).  \label{eq:HImag}
\end{equation}

\subsubsection{Dynamics}
\label{app:realDynamics}

Here we briefly review some dynamical properties of real-valued quantum mechanics.
We want $iH$ to be real such that no imaginary values appear in Schr\"{o}dinger's equation,
  which implies that $H$ must be imaginary and antisymmetric.
Thus for each eigenvector $\ket{E}$,
  taking the complex-conjugate of $H\ket{E}=E\ket{E}$ (in any basis) implies that
  the complex-conjugated $\ket{E}^*$ has eigenvalue $-E$.
Therefore, eigenvectors with nonzero eigenvalues come in complex-conjugate pairs with negated eigenvalues.
It is convenient to consider real-valued linear combinations
  $\ket{E}_\pm = \tfrac{1}{\sqrt{\pm2}} (\ket{E} \pm \ket{-E})$
  for each $E>0$.
The coefficients $a_{E,\pm}(t)$ of a wavefunction
  $\ket{\psi(t)} = \sum_{E,\pm} a_{E,\pm}(t) \ket{E}_\pm$
  obey the time evolution $\partial_t a_{E,\pm}(t) = \mp E a_{E,\mp}(t)$ for each $E>0$.

A real-valued wavefunction can not be in an eigenstate $\ket{E}$ with nonzero $E$ since $\ket{E}$ is complex-valued when $E\neq0$.
In complex-valued quantum mechanics, eigenstates are steady-states,
  i.e. states that do not change with time.
The analog of steady-states in real-valued quantum mechanics
  are oscillating superpositions of $\ket{E}_\pm$:
\begin{equation}
\begin{aligned}
  \ket{\psi(t)} &= \cos(Et) \ket{E}_+ - \sin(Et) \ket{E}_- \\
  &= \tfrac{1}{\sqrt{2}} e^{-i E t} \ket{E}
   + \tfrac{1}{\sqrt{2}} e^{+i E t} \ket{-E}.
\end{aligned}
\end{equation}
The analog of the lowest-energy ground state is the above state with the largest possible $E$.
Notably, this ``steady-state'' looks like a cat state from the perspective of complex-valued quantum mechanics.
Indeed, the steady-state of two disconnected subsystems will share an entangled qubit
  from the perspective of complex-valued quantum mechanics.
For example, if $H = - Y_1 - Y_2$ is the Hamiltonian for two qubits,
  then the lowest-energy steady-state is not a tensor product state;
  instead, it is maximally entangled:
\begin{equation}
\begin{aligned}
  \ket{\psi(t)}
  &= \big(\cos(t)\ket{\uparrow} - \sin(t)\ket{\downarrow} \big) \otimes
     \big(\cos(t)\ket{\uparrow} - \sin(t)\ket{\downarrow} \big) \\
  &\,- \big(\sin(t)\ket{\uparrow} + \cos(t)\ket{\downarrow} \big) \otimes
     \big(\sin(t)\ket{\uparrow} + \cos(t)\ket{\downarrow} \big). \label{eq:realGround}
\end{aligned}
\end{equation}

In complex-valued quantum mechanics, the following tensor product axiom holds for combining quantum states of two systems:
  ``the state representing two independent preparations of the two systems is the tensor product of the two preparations'' \cite{RealQMFalsifiable}.
However, this axiom does not hold in real-valued quantum mechanics.
Indeed, \refcite{RealQMExperiment,RealQMExp} have verified that real-valued quantum mechanics with this tensor product axiom is not consistent with experiment.
Nevertheless,
  our analysis above shows that real-valued quantum mechanics is consistent with complex-valued quantum mechanics if the tensor product axiom is dropped.
Indeed the tensor product axiom is not consistent with our local mapping \eqref{eq:localWavefunctionMap}, which maps complex-valued tensor product wavefunctions to a sum of two tensor products.
Furthermore, \eqnref{eq:realGround} shows that a sum of two tensor products arises naturally for ground state wavefunctions of two disconnected subsystems in real-valued quantum mechanics.

\subsection{Zero Sum}
\label{app:zero}
To satisfy the zero-sum conditions (\ref{eq:sumPsi}--\ref{eq:sumH0}), we once again introduce an additional qubit.
We map the wavefunction $\ket{\psi}$ to
\begin{equation}
    \ket{\widetilde{\psi}} = \ket{\psi} \otimes \ket{-}
\end{equation}
where 
\begin{equation}
    |-\rangle = \tfrac{1}{\sqrt{2}}\big(\ket{\uparrow}-\ket{\downarrow} \big).
\end{equation}
Similarly, any operator $Q$ can be linearly mapped to a new operator
\begin{equation}
    \widetilde{Q} = Q\otimes |-\rangle\langle -|,
\end{equation}
where
\begin{equation}
    |-\rangle\langle -| = \tfrac{1}{2}\big(\ket{\uparrow}\bra{\uparrow}-\ket{\uparrow}\bra{\downarrow}-\ket{\downarrow}\bra{\uparrow}+\ket{\downarrow}\bra{\downarrow}\big).
\end{equation}
By construction, $\ket{\widetilde{\psi}}$ and $\widetilde{Q}$ obey the zero-sum conditions:
\begin{equation}
\begin{aligned}
    \sum_i \widetilde{\psi}_i &= 0 \\
    \sum_i \widetilde{Q}_{ij} &=0= \sum_j \widetilde{Q}_{ij}.
\end{aligned}
\end{equation}
  in any basis where the additional qubit has basis vectors $\ket{\uparrow}$ and $\ket{\downarrow}$.
Furthermore, the zero-sum constraints are preserved under evolution under the Schr\"{o}dinger equation for a Hamiltonian that obeys the zero-sum constraints.

In order to preserve geometric locality,
  we add another set of new qubits adjacent to each $x$.
The mapping for wavefunctions and local operators (including terms in the Hamiltonian) is
\begin{align}
  \ket{\psi} &\to \ket{\widetilde{\psi}} = \ket{\psi} \otimes_x \ket{-}_x \\
  Q_x &\to \widetilde{Q}_x = Q_x \otimes \ket{-}_x \bra{-}_x. \label{eq:Qx0Sum}
\end{align}

\subsection{Examples}

\subsubsection{Ising (systematic mapping)}

As an example, we can consider applying these systematic mappings to a transverse-field Ising model Hamiltonian $H=\sum_x H_x$:
\begin{align}
  H_x &= - J Z_x Z_{x+1} - h Y_x \nonumber\\
      &\to J Z_x Z_{x+1} \otimes \sigma^2_x - h Y_x \\
      &\to \left( J Z_x Z_{x+1} \otimes \sigma^2_x - h Y_x \right) \otimes \ket{-}_x\bra{-}_x. \nonumber
\end{align}
$X_x$, $Y_x$, $Z_x$ are Pauli operators in the Ising model Hamiltonian,
  while $\sigma_x^2$ is a Pauli operator acting on additional qubits.
$\ket{-}_x\bra{-}_x$ acts on an additional set of qubits.
A transverse $h Y_x$ term was considered in the first line instead of the traditional $h X_x$ term to make the example more useful.
The second line is the result of applying \eqnref{eq:HImag}
  to obtain an imaginary-valued Hamiltonian.
\eqnref{eq:Qx0Sum} is applied to obtain the third line,
  for which the resulting Hamiltonian has zero row and column sums.

\subsubsection{Ising (clever duality)}

However, adding additional qubits is not always necessary.
For example, the XY Hamiltonian
  is dual to an imaginary-valued Hamiltonian with zero row and column sums:
\begin{align}
  H_x &= -J_{XY} \, (X_x X_{x+1} + Y_x Y_{x+1}) \label{eq:XXZ}\\
      &\leftrightarrow J_\text{XY} \, (X_{x-1} Y_x X_{x+1} - Y_x)
\end{align}
The second Hamiltonian describes the phase transition between a SPT cluster state \cite{ClusterSPT} and the trivial disorered phase.

The duality mapping used above is:
\begin{equation}
\begin{aligned}
  Y_x Y_{x+1} &\leftrightarrow Y_x \\
  Z_x &\leftrightarrow X_{x-1} X_x
\end{aligned}
\end{equation}
which implies that $X_x X_{x+1} \leftrightarrow -X_{x-1} Y_x X_{x+1}$.
This duality only maps symmetric operators to local operators.
Symmetric operators are operators that commute with certain $Z_2$ symmetries,
  which are $\prod_x Z_x$ and $\prod_x Y_x$ for the respective left and right sides of the duality.
The duality maps the $Z_2$ symmetries to the identity (ignoring boundary conditions):
\begin{equation}
\begin{aligned}
  \prod_x Z_x &\leftrightarrow 1 \\
  1 &\leftrightarrow \prod_x Y_x
\end{aligned}
\end{equation}
These properties are common in duality mappings,
  such as the self-duality \cite{RadicevicDuality} of the transverse-field Ising model.

\section{Emergent Lorentz Invariance}
\label{app:Lorentz}

In \secref{sec:fast}, we had to posit that the observed Lorentz invariance in our universe is emergent (rather than exact).
This may not be a major hurdle since emergent Lorentz invariance has been shown to be
  a stable fixed point under the renormalization group (RG) in several strongly-coupled models
  \cite{RoyLI,BednikLI,BelenchiaLI}.
Indeed, emergent Lorentz invariance is rather ubiquitous in low-energy physics.
For example, emergent Lorentz invariance occurs in materials such as graphene,
  which exhibits an electron band structure with a Lorentz-invariant Dirac cone at low energy.
As a result, the electrons in graphene experience a Lorentz-invariant speed limit that is much smaller than
  the speed of photons.

But with multiple species of fermions,
  emergent Lorentz invariance would require that all Dirac fermions have the same velocity.
Although the velocities for different species flow to the same value under RG,
  the flow is very slow; hence emergent Lorentz invariance seems to require fine tuning in order to be consistent with certain very precise experiments. \cite{LorentzFineTuning,PolchinskiLorentz}
Furthermore, general relativity would likely also need to be emergent in this scenario \cite{LaughlinEmergent,emergentGravityCallenges}.

It is not clear if including the effects of a fully quantized theory of emergent quantum gravity could significantly affect the RG flow such that Lorentz invariance emerges more rapidly.
In particular, we note that when Lorentz invariance is broken,
  the local diffeomorphism ``gauge symmetry'' of general relativity is broken.
This reminds us of $(3+1)$-dimensional U(1) gauge theory,
  in which we could imagine adding an $A^2 = A_\mu A^\mu$ term to the Lagrangian, which breaks the gauge symmetry.
However, although $A^2$ naively appears to be relevant in $(3+1)$ dimensions since $A^2$ has energy dimension 2,
  $A^2$ is actually an irrelevant perturbation.
In fact, all perturbations to $(3+1)$-dimensional U(1) gauge theory are irrelevant, including the perturbations that break gauge invariance \cite{stableU1}.

We emphasize that although these gauge-symmetry-breaking terms are irrelevant for the compact gauge group U(1),
  such terms are \emph{relevant} (as naively expected) for the non-compact gauge group $\mathbb{R}$.
Note that both of these gauge groups lead to the same classical equations of motion since they have the same Lie algebra.
It is remarkable that it is only after quantizing these two different gauge theories, with gauge groups U(1) or $\mathbb{R}$,
  that we discover that gauge invariance can be emergent in U(1) gauge theory but not in $\mathbb{R}$.
With this in mind, we speculate that
emergent Lorentz invariance may occur sufficiently rapidly in some theories of emergent gravity such that it could be consistent with experiments. \cite{WenGravity}

One approach to intuitively understanding why $A^2$ is irrelevant in U(1) gauge theory
  is to consider the corresponding lattice gauge theory (without constraining the Hilbert space to the gauge-invariant states).
The lattice gauge theory is similar to the toric code \cite{toricCode},
  except the $Z_2$ qubits are replaced by integer-valued degrees of freedom $E_e \in \mathbb{Z}$ on each edge $e$ of the lattice.
The Hamiltonian is
  $H = \sum_i (\nabla_i n)^2 - \sum_p \cos(\nabla_p \times A)$.
$\nabla_i n$ denotes the lattice divergence of $n_e$ centered at the vertex $i$,
  while $\nabla_p \times A$ is the lattice curl of $A_e$ around the plaquette $p$.
The commutation relations are $[e^{i A_e}, n_{e'}] = \delta_{e,e'}$.
The states with $\nabla_i n = 0$ obey Gauss's law and are the gauge-invariant states.

Gauge invariance is often imposed by taking $\nabla_i n = 0$ to be a constraint on the Hilbert space.
Here, we do not constrain the Hilbert space;
  instead we use the first term of the Hamiltonian to impose an energy penalty on states that are not gauge invariant.
Since $n_e$ is integer valued,
  all energy excitations of $\sum_i (\nabla_i n)^2$ cost \emph{finite energy},
  which makes the gauge invariance stable to perturbations.
For example, a $\epsilon A^2$ term in the U(1) gauge theory Lagrangian is analogous to a
  $\epsilon \sum_e \cos(A_e)$ term in the lattice gauge theory Hamiltonian.
If $\epsilon$ is sufficiently small,
  this term does not does not lead to confinement
  since its excitations cost finite energy due to the $(\nabla_i n)^2$ term.
This can be shown more formally using degenerate perturbation theory \cite{SchriefferWolff}.
Similarly, in the toric code (i.e. $Z_2$ lattice gauge theory where Gauss's law enters the Hamiltonian as an energy penalty rather than a Hilbert space constraint),
  arbitrary perturbations are irrelevant and do not destabilize the emergent gauge invariance or topological order. \cite{toricCode}

We emphasize that even if there are faster-than-light degrees of freedom,
  it could still be very difficult for observers to send information faster than light.
Suppose a Hamiltonian $H$ is Lorentz-invariant
  (either exactly or approximately at low energies) with velocity $c$,
  and suppose an observer living within the wavefunction $\Psi$ wants to use the fast classical degrees of freedom
  to send signals faster than $c$
  (without exceeding low energies if the Lorentz-invariance is approximate).
If $\Psi(t)$ is well-described by Schr\"{o}dinger's equation for a local Hamiltonian,
  then the observer will only be able to send signals much faster than $c$
  by taking advantage of very small possible violations of Schr\"{o}dinger's equation \cite{PolchinskiNonlinearEPR,GisinNonlinearEPR}.

\section{Approximate Simulation}
\label{app:approximate}


As noted in \eqnref{eq:CPU},
  simulating the EmQM model is very CPU intensive.
Fortunately,
  we can approximately simulate the EmQM model significantly faster by ``integrating out'' the fast degrees of freedom (i.e. the forward and backward-propagating bits) 
  so that we only have to directly simulate the slow degrees of freedom (i.e. the stochastic matrices).
This will allow us to simulate $\Delta_\text{jump} \gg S$ many time steps all at once with negligible error.

To do this,
  we approximate the stochastic matrices $M_{s,x}$ as constant over $\Delta_\text{jump}$ many time steps.
We then estimate how much the stochastic matrices might change after these $\Delta_\text{jump}$ time steps.
If $\Delta_\text{jump} \gg N = 2^n$,
  then each bit string will occur many times over this many time steps.
Therefore, simulating the propagation of $\Delta_\text{jump}$ many bit strings involves a lot of duplication of effort.
To speed up the simulations,
  we can instead just calculate how much each stochastic matrix will be affected by the propagation of all $N$ possible bit strings,
  each weighted by its probability times $\Delta_\text{jump}$.

\subsection{Approximate Algorithm}

The approximate algorithm to compute how $\Delta_\text{jump}$ time steps could affect the stochastic matrices is as follows:

We first compute $\bm{P}_S^{(\tau)}$ using \eqnref{eq:Ps} to get the probability vector for the output bits $a_S$ of the stochastic circuit.

Then for each $\gamma=1,2$,
  we sample an $N \times N$ matrix $\beta_{S,\gamma}^{(\tau)}$
  from the multinomial distribution of $\Delta_\text{jump}$ trials with a matrix of probabilities $(B^{(+\gamma)} \cdot \bm{P}_S^{(\tau)}) \otimes (B^{(-\gamma)} \cdot \bm{P}_S^{(\tau)})$,
  where $B^{(\gamma)}$ is the $N\times N$ stochastic matrix defined in \eqnref{eq:B}.
Therefore, $\beta_{S,\gamma}^{(\tau)}(b_{S,+\gamma},b_{S,-\gamma})$ counts how many times the pair $(b_{S,+\gamma},b_{S,-\gamma})$ of bit strings could have occurred
  after $\Delta_\text{jump}$ time steps.
Note that we approximated $M_{s,x}^{(\tau)}$ as constant over these $\Delta_\text{jump}$ time steps.

Next we back-propagate [analogous to \eqnref{eq:Qb}] $\beta_{s,\gamma}^{(\tau)}$ using the permutation matrices $Q_s$ defined in \eqnref{eq:Qs}:
\begin{equation}
  \beta_{s,\gamma}^{(\tau)} = Q_s^T \cdot \beta_{s+1,\gamma}^{(\tau)} \cdot Q_s
\end{equation}

Finally, we wish to update the perturbations $m_{s,x}$ in accordance with \eqnref{eq:dm} using $\beta_{S,\gamma}^{(\tau)}$.
To do this, we modify \eqnref{eq:dm} to
\begin{align}
  &m_{s,x}^{(\tau+\Delta_\text{jump})} = m_{s,x}^{(\tau)} + \Delta_\text{m}
     \sum_{b_\xx^+} \sum_{b_\xx^-} \sum_{e_\xx} \Bigg[ \label{eq:dmNew}\\ 
    &\quad\quad \sum_{\gamma=1,2} \beta_{s,x,\gamma}^{(\tau)}(b_\xx^+,b_\xx^-) \, p_{s,x}^{(\tau)}(e_\xx|b_\xx^-)
    \left( \bm{\hat{b}}_\xx^+ - \bm{\hat{b}}_\xx^- \right) \otimes \bm{\hat{e}}_\xx \Bigg] \nonumber
\end{align}
  where $b_\xx^+$, $b_\xx^-$, and $e_\xx$ are each a pair of bits (00, 01, 10, or 11).
These bit pairs index the $4\times4$ matrices $m_{s,x}^{(\tau)}$ and $\beta_{s,x,\gamma}^{(\tau)}$.
The bit pairs also determine the basis 4-vectors $\bm{\hat{b}}_\xx^\pm$ and $\bm{\hat{e}}_\xx$;
  e.g. $\bm{\hat{b}}_\xx^+ = (1,0,0,0)$ if $b_\xx^+ = 00$.
Above, $\beta_{s,x,\gamma}^{(\tau)}(b_\xx^+,b_\xx^-)$ counts how many times the pair $(b_\xx^+,b_\xx^-)$ could have occurred over the $\Delta_\text{jump}$ time steps.
That is,
\begin{equation}
  \beta_{s,x,\gamma}^{(\tau)} = \sum_{b^+} \sum_{b^-} \bm{\hat{b}}_\xx^+ \otimes \bm{\hat{b}}_\xx^- \, \beta_{s,\gamma}^{(\tau)}(b^+,b^-)
\end{equation}
  where $\sum_{b^\pm}$ sums over the $N=2^n$ different bit strings of $n$ bits, and
  $\bm{\hat{b}}_\xx^\pm$ is the basis 4-vector that depends on bits $x$ and $x+1$ of the bit string $b^\pm$;
  e.g. $\bm{\hat{b}}_\xx^\pm = (1,0,0,0)$ [or $(0,1,0,0)$] if bits $x$ and $x+1$ of $b^\pm$ are $00$ [or $01$].
In \eqnref{eq:dmNew},
  $p_{s,x}^{(\tau)}(e_\xx|b_\xx^-)$ is a conditional probability distribution for $e_\xx$ given $b_\xx^-$.

We need to calculate $p_{s,x}^{(\tau)}(e_\xx|b_\xx^-)$ such that it is approximately equal to the probability that $\bm{\hat{e}}_{s-1,x,\gamma}^{(\tau)} = \bm{\hat{e}}_\xx$ given $\bm{\hat{b}}_{s,\xx,-\gamma}^{(\tau)}=\bm{\hat{b}}_\xx^-$ in \eqnref{eq:dm} over the $\Delta_\text{jump}$ time steps.
Recall that $\bm{\hat{e}}_{s-1,x,\gamma}^{(\tau)}$ is chosen uniformly at random from the set of basis 4-vectors that keep $M_{s,x}^{(\tau)} = Q_{s,x} + m_{s,x}^{(\tau)}$ non-negative.
Therefore we can take $p_{s,x}^{(\tau)}(e_\xx|b_\xx^-) = 1/4$ to be uniform probabilities
  as long as this results in a stochastic matrix $M_{s,x}^{(\tau+\Delta_\text{jump})}$ with non-negative entries.

However if $M_{s,x}^{(\tau+\Delta_\text{jump})}$ has negative entries,
  then $p_{s,x}^{(\tau)}$ can not be uniform.
Instead, for each negative $M_{s,x}^{(\tau+\Delta_\text{jump})}(b_\xx^-,e_\xx)<0$ entry that we find,
  we must make $p_{s,x}^{(\tau)}(e_\xx|b_\xx^-)$ a free parameter
  [while uniformly adjusting the other entries such that $p_{s,x}^{(\tau)}(e_\xx|b_\xx^-)$ is a probability distribution for $e_\xx$ given $b_\xx^-$].
We then solve for these free parameters such that the previously-negative entries of $M_{s,x}^{(\tau+\Delta_\text{jump})}$ are zero.
If $M_{s,x}^{(\tau+\Delta_\text{jump})}$ for the new $p_{s,x}^{(\tau)}$ has additional negative entries,
  then we repeat the procedure until all entries are non-negative.
That is, we make more entries in $p_{s,x}^{(\tau)}$ free parameters and re-solve for the old and new free parameters such that any previously-negative entry of $M_{s,x}^{(\tau+\Delta_\text{jump})}$ is zero.
At most, $p_{s,x}^{(\tau)}$ can only have $4\times3$ free parameters,
  since it describes 4 different probability distributions, each with up to 3 free parameters.
Therefore, the above procedure can not repeat more than 12 times.
As desired, this procedure results in a $p_{s,x}^{(\tau)}$ that is
  approximately equal to the probability that $\bm{\hat{e}}_{s-1,x,\gamma}^{(\tau)} = \bm{\hat{e}}_\xx$ given $\bm{\hat{b}}_{s,\xx,-\gamma}^{(\tau)}=\bm{\hat{b}}_\xx^-$ in \eqnref{eq:dm} over the $\Delta_\text{jump}$ time steps.

\subsection{Approximation Error}

The approximate algorithm makes the approximation that $M_{s,x}^{(\tau)}$ is constant over $\Delta_\text{jump}$ many time steps.
This approximation is very similar to the cause of the deviation $\varepsilon_\text{delay}(t)$ [\eqnref{eq:epsLag}]
  that results from the $O(S)$ discrete time delay for the classical bits to move through the circuit.
The primary difference is that the delay is $\Delta_\text{jump}$ instead of $O(S)$ for the approximate algorithm.
Therefore, the approximate algorithm results in an error similar to \eqnref{eq:epsLag},
  except a factor of $S$ is replaced by $\Delta_\text{jump}$:
\begin{equation}
\begin{aligned}
  \varepsilon_\text{jump}(t) &\sim \frac{\Delta_\text{jump}}{S} \varepsilon_\text{delay}(t) \\
  &\sim n \Delta_t \Delta_\text{jump} \, t
\end{aligned} \label{eq:eps approx}
\end{equation}

\begin{figure}[t!]
  \centering
  \includegraphics[width=\columnwidth]{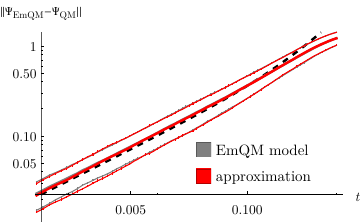}
  \caption{%
    The deviation $||\Psi(t) - \Psi_\text{QM}(t)||$ vs time $t$ for $n=4$ qubits with $\epsilon_0=1$, showing agreement between
      the EmQM model (gray), approximate algorithm with $\epsilon_\text{j} = 0.02$ (red), and estimated $\varepsilon(t)$ from \eqnref{eq:err} (dashed black).
    For the EmQM model and approximate algorithm,
        we average over 1000 random realizations and plot the mean (thick lines)
        along with the mean plus/minus the standard deviation (thin lines)
        to demonstrate that both the mean and statistical fluctuations agree well.
    Error bars denote one standard deviation of statistical error resulting from the finite number of 1000 samples.
  }\label{fig:validateFast}
\end{figure}

\begin{figure}[t!]
  \centering
  \includegraphics[width=\columnwidth]{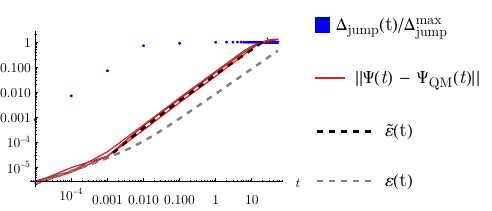}
  \caption{%
    The deviation $||\Psi(t) - \Psi_\text{QM}(t)||$ (red) vs time $t$ calculated using the approximate algorithm with a large $\varepsilon_\text{jump} = 10$
      so that we can study errors due to the approximate simulation algorithm.
    Simulations use $n=4$ qubits and three different random initializations
      with $\varepsilon = 0.01$ [in \eqnref{eq:epsilon}].
    In all simulations in this work, $\Delta_\text{jump}$ is chosen to be as large as possible without exceeding $\Delta_\text{jump}^\text{max}$.
    $\Delta_\text{jump}$ is sometimes limited (shown by blue dots)
      when the time step is very small (due to the logarithmic time axis).
    By inserting the actual time-dependent $\Delta_\text{jump}(t)$ into \eqnref{eq:eps approx},
      our estimate
      $\widetilde{\varepsilon}(t)$ [\eqref{eq:epsTilde}, dashed black line]
      for the deviation (under the influence of the approximation)
      matches the $\varepsilon_\text{jump}(t)$-dominated
      simulation data (red) remarkably well.
    In contrast, the dashed gray curve shows the expected deviation $\varepsilon(t)$ of the EmQM model without approximation,
      which is significantly smaller when $0.01 \lesssim t \lesssim 10$.
    This validates \eqnref{eq:epsTilde}.
  }\label{fig:fidelityFastError}
\end{figure}

In our simulations, we want to choose $\Delta_\text{jump}$ to be as large as possible
  without introducing noticeable errors in our plots.
To achieve this, we choose $\Delta_\text{jump}$ such that $\varepsilon_\text{jump}(t)$ in
  \eqnref{eq:eps approx} is parameterized by a new control parameter $\epsilon_\text{j}$:
\begin{equation}
  \varepsilon_\text{jump}(t) \sim \epsilon_0 \, \epsilon_\text{j} \, t
\end{equation}
Small $\epsilon_\text{j}$ will then ensure that $\varepsilon_\text{jump}(t)$ contributes negligibly to $\varepsilon(t) = ||\Psi - \Psi_\text{QM}||$ in comparison to the estimated $\varepsilon(t) \sim \sqrt{\epsilon_0 t}$ in \eqnref{eq:err}.
Therefore, \eqnref{eq:eps approx} implies that we must limit $\Delta_\text{jump}$ to be no larger than
\begin{equation}
\begin{aligned}
  \Delta^\text{max}_\text{jump}
  &\approx \frac{\epsilon_0 \, \epsilon_\text{j}}{\Delta_t n} \\
  &\sim \frac{n^3 N^2}{\epsilon_0^5} \epsilon_\text{j}.
\end{aligned}
\end{equation}
Due to this additional source of error,
  the approximate algorithm will produce slightly larger deviations from quantum mechanics,
  which we estimate to be
\begin{equation}
  \widetilde{\varepsilon}(t) \sim \sqrt{\varepsilon(t)^2 + \varepsilon_\text{jump}(t)^2}. \label{eq:epsTilde}
\end{equation}

To ensure that the error $\varepsilon_\text{jump}(t)$ from the approximate algorithm remains negligible,
  we use small $\epsilon_\text{j} = 0.02$ for most simulations (excluding \figref{fig:fidelityFastError}).
The only exception is the $\epsilon_0 = 0.05$ data in \sfigref{fig:fidelity}{b,d},
  which used $\epsilon_\text{j} = 0.1$ in order to keep the simulation time under one month.

To validate that our approximate algorithm does not significantly affect the deviation $||\Psi - \Psi_\text{QM}||$ or its statistical fluctuations when $\varepsilon_\text{jump}$ is small,
  in \figref{fig:validateFast} we compare these quantities for the EmQM model and the approximate algorithm.
To validate \eqnref{eq:epsTilde},
  in \figref{fig:fidelityFastError} we plot the deviation $||\Psi(t) - \Psi_\text{QM}(t)||$ from quantum mechanics using the approximate model with large $\epsilon_\text{j}$
  such that the error from approximation dominates.

\end{document}